\documentclass[prb,amsmath,amssymb,
twocolumn,
floatfix,
superscriptaddress]{revtex4-1}
\usepackage{siunitx}
\usepackage{graphicx}
\usepackage[colorlinks]{hyperref}
\usepackage{tikz}
\usepackage{braket}

\makeatletter\def\l@section#1#2{}\makeatother 

\definecolor{linkcolor}{RGB}{0,83,166}
\hypersetup{
  colorlinks = true,
  allcolors = {linkcolor}
}

\newcommand{\dwaddress}{D-Wave Systems, Burnaby, British Columbia, Canada}

\begin{document}
\title{
  Scaling advantage in quantum simulation of geometrically frustrated magnets
}

\author{Andrew~D.~King}\email[]{aking@dwavesys.com}
\author{Jack~Raymond}
\thanks{These authors contributed equally.}
\author{Trevor~Lanting}
\thanks{These authors contributed equally.}
\affiliation{\dwaddress}
\author{Sergei~V.~Isakov}
\affiliation{Google, 8002 Zurich, Switzerland}
\author{Masoud~Mohseni}
\affiliation{Google, Venice, California 90291, USA}
\author{Gabriel~Poulin-Lamarre}
\affiliation{\dwaddress}
\author{Sara Ejtemaee}
\author{William Bernoudy}
\author{Isil Ozfidan}
\author{Anatoly~Yu.~Smirnov}
\author{Mauricio~Reis}
\author{Fabio~Altomare}
\author{Michael Babcock}
\author{Catia Baron}
\author{Andrew~J.~Berkley}
\author{Kelly Boothby}
\author{Paul~I.~Bunyk}
\author{Holly~Christiani}
\author{Colin Enderud}
\author{Bram Evert}
\author{Richard Harris}
\author{Emile Hoskinson}
\author{Shuiyuan Huang}
\author{Kais~Jooya}
\author{Ali~Khodabandelou}
\author{Nicolas Ladizinsky}
\author{Ryan Li}
\author{P.~Aaron Lott}
\author{Allison~J.~R.~MacDonald}
\author{Danica~Marsden}
\author{Gaelen Marsden}
\author{Teresa Medina}
\author{Reza Molavi}
\author{Richard Neufeld}
\author{Mana Norouzpour}
\author{Travis~Oh}
\author{Igor Pavlov}
\author{Ilya Perminov}
\author{Thomas Prescott}
\author{Chris Rich}
\author{Yuki Sato}
\author{Benjamin Sheldan}
\author{George~Sterling}
\author{Loren J.~Swenson}
\author{Nicholas Tsai}
\author{Mark~H.~Volkmann}
\author{Jed~D.~Whittaker}
\author{Warren Wilkinson}
\author{Jason Yao}
\affiliation{\dwaddress}
\author{Hartmut Neven}
\affiliation{Google, Venice, California 90291, USA}
\author{Jeremy~P.~Hilton}
\affiliation{\dwaddress}
\author{Eric Ladizinsky}
\affiliation{\dwaddress}
\author{Mark W.~Johnson}
\affiliation{\dwaddress}
\author{Mohammad H.~Amin}
\affiliation{\dwaddress}
\affiliation{Department of Physics, Simon Fraser University}
\date{\today}

\begin{abstract}
  The promise of quantum computing lies in harnessing programmable quantum devices for practical applications such as efficient simulation of quantum materials and condensed matter systems.
  One important task is the simulation of geometrically frustrated magnets\cite{King2018,Isakov2003,Kamiya2012,Li2019} in which topological phenomena can emerge from competition between quantum and thermal fluctuations.
  Here we report on experimental observations of relaxation in such simulations, measured on up to 1440 qubits with microsecond resolution.
  By initializing the system in a state with topological obstruction, we observe quantum annealing (QA) relaxation timescales in excess of one microsecond.  
  Measurements indicate a dynamical advantage in the quantum simulation over the classical approach of path-integral Monte Carlo (PIMC) fixed-Hamiltonian relaxation with multiqubit cluster updates.
  The advantage increases with both system size and inverse temperature, exceeding a million-fold speedup over a CPU.
  This is an important piece of experimental evidence that in general, PIMC does not mimic QA dynamics for stoquastic Hamiltonians.
  The observed scaling advantage, for simulation of frustrated magnetism in quantum condensed matter, demonstrates that near-term quantum devices can be used to accelerate computational tasks of practical relevance.
\end{abstract}

\maketitle

\begin{figure*}
  \includegraphics[scale = .77]{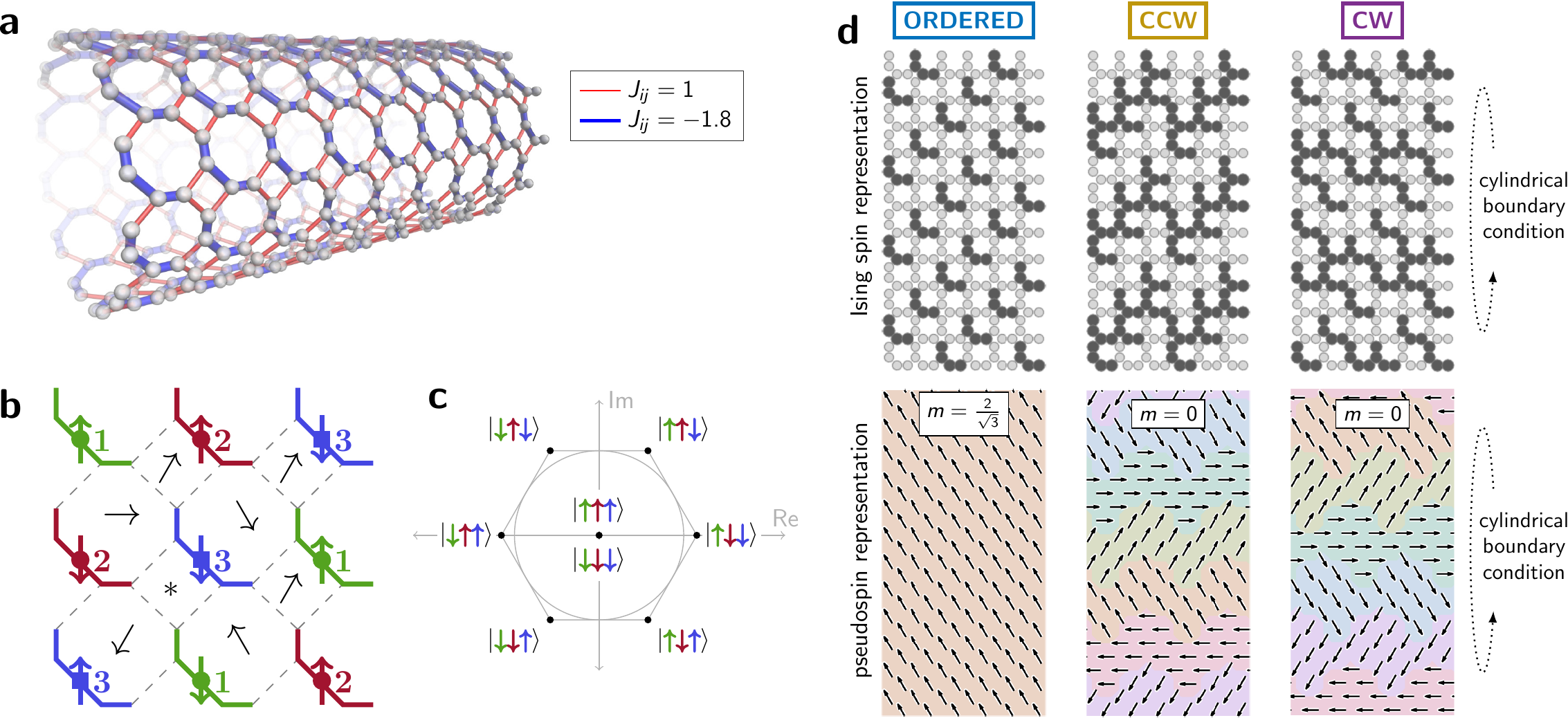}
  \caption{{\bf Geometrically frustrated lattice and escape from topological obstructions.}
    {\bf a}, We program the superconducting processor to simulate a lattice with cylindrical boundary conditions. Each square or octagonal plaquette is frustrated, having an odd number of antiferromagnetic (red) couplers that cannot be satisfied simultaneously.
    {\bf b--c}, Each plaquette is composed of spins in three sublattices (indicated by green, red, blue) ({\bf b}).  Due to frustration, each plaquette has six ground states, which map to six orientations of a {\em pseudospin} based on the value of spins in the three incident sublattices ({\bf c}).  Adding quantum fluctuations leads to ferromagnetic order in the pseudospins.
    {\bf d}, We initialize the simulation in three initial conditions: an ordered state, and states in which the pseudospin winds around the periodic dimension (top/bottom) counterclockwise (CCW) or clockwise (CW).  Ordered states maximize the order parameter $m$, while CCW and CW have $m=0$.  Spin and pseudospin interpretations are shown.  We simulate relaxation from these initial conditions using quantum (QA) and classical (PIMC) methods.
  }
  \label{fig:1}
\end{figure*}

The recent demonstration of quantum computational supremacy in a superconducting processor\cite{Arute2019} lends credence to the ultimate viability of quantum computing, but leaves open the question of quantum advantage in practical applications.  In absence of quantum error correction, the utility of gate-model quantum computers has yet to be determined \cite{Preskill2018}. 
Meanwhile, analog quantum simulations are an attractive proving ground for near-term quantum devices: the application can be chosen to suit the strengths of the platform, allowing the demonstration of a variety of engineered many-body quantum phenomena in noisy hardware.  Examples include the Kibble-Zurek mechanism in Rydberg atoms\cite{Keesling2019}, dynamical phase transitions in trapped ions\cite{Zhang2017b,Jurcevic2017}, localization in superconducting transmon qubits\cite{Roushan2017}, Coulomb blockade in quantum dots\cite{Hensgens2017}, and phase transitions in superconducting flux qubits\cite{Harris2018,King2018}.  Having established the possibility of simulating complex quantum phenomena on a manufactured quantum device, as Feynman famously proposed\cite{Feynman1982}, one comes to the next question: can the quantum device confer a computational advantage?

We address this question using a superconducting flux-qubit quantum annealing (QA) processor.  In this case---although not in general\cite{Ozfidan2019}---the simulated system is described by a {\em stoquastic} Hamiltonian\cite{Bravyi2008}: its equilibrium properties can be studied using path-integral Monte Carlo (PIMC)\cite{Suzuki1976} with no {\em sign problem}\cite{Troyer2005}.  Here the memory requirements of PIMC are polynomial in system size  for a given nonzero temperature.  However, relaxation dynamics can become exponentially slow in both QA and PIMC.  It is therefore natural to ask whether QA can estimate equilibrium statistics faster than PIMC.

This question has been the subject of controversy\cite{Heim2015,Isakov2016,Andriyash2017,Kechedzhi2018,Mbeng2018}, and was recently studied in two specific settings for stoquastic Hamiltonians.  For simple systems, with few dominating homotopy-inequivalent tunneling paths, the scaling of incoherent QA and PIMC is the same\cite{Isakov2016}.  However, this equivalence is not general, and does not apply to complex (highly frustrated) systems with many tunneling paths and coherent dynamics\cite{Andriyash2017,Kechedzhi2018}.  Here we provide the first experimental demonstration that even in the presence of noise and finite temperature, a QA processor can provide a scaling advantage---in both size and temperature---over PIMC relaxation dynamics for a 2-local stoquastic Hamiltonian.  As the simulated system becomes larger and colder, QA shows a growing advantage over PIMC in relaxation timescales, reaching a maximum factor of $3\times 10^6$.

We demonstrate this speedup in the dynamics of a {\em geometrically frustrated} lattice\cite{Moessner2001}.  In such a system, the lattice geometry is a repeating pattern of plaquettes, all of which are {\em frustrated}: each plaquette contains competing terms that cannot be satisfied simultaneously.
It was recently shown that the exotic physics of a topological phase transition in this type of frustrated system\cite{Isakov2003} can be simulated in a QA processor \cite{King2018}.  However, the methods and apparatus did not allow accurate measurement of relaxation timescales.  Here we introduce new experimental methods and apply them using an improved QA processor fabricated with lower-noise processes (see Methods).  We realize a programmable transverse field Ising model (TFIM), whose Hamiltonian can be written as
\begin{equation}\label{eq:ham}
  H = H(s) = J(s)\sum_{i<j}J_{ij}\sigma_i^z\sigma_j^z - \Gamma(s)\sum_{i}\sigma_i^x
  \end{equation}
  where $s$ is an annealing parameter, $J_{ij}$ are 2-local coupling terms, $J(s)$ is a universal coupling energy prefactor, $\sigma_i^z$ and $\sigma_i^x$ represent Pauli operators on the $i$th spin, and $\Gamma(s)$ is the transverse field, which induces quantum fluctuations.  We measure system evolution under fixed $H(s)$; this is a departure from the traditional QA use case, where $H(s)$ varies smoothly with time.

  We program the couplings $J_{ij}$ of the QA processor with a square-octagonal lattice (Fig.~\ref{fig:1}a).  This lattice exhibits {\em order-by-disorder}, a phenomenon in which the addition of quantum fluctuations---a source of dynamical disorder---drastically changes the low-temperature properties of the system, creating long-range order that does not exist in the classical setting\cite{Wannier1950}.  Due to geometric frustration in the lattice, each four- or eight-spin plaquette has six classical ground states; these can be represented by six orientations of a {\em pseudospin} (Fig.~\ref{fig:1}b-c).  A combination of these local ground states can form a global ground state provided that neighboring pseudospins are approximately aligned, but this is not enough to impose long-range order in the classical case.  The addition of a transverse field energetically favors ``flippable'' four-qubit chains, which assume symmetric Greenberger-Horne-Zeilinger (GHZ) superposition; maximizing the number of flippable chains forces the pseudospins to align ferromagnetically.  The magnitude $m$ of the average pseudospin quantifies this order, which varies as a function of transverse field and temperature.

\begin{figure*}
  \includegraphics[scale = .775]{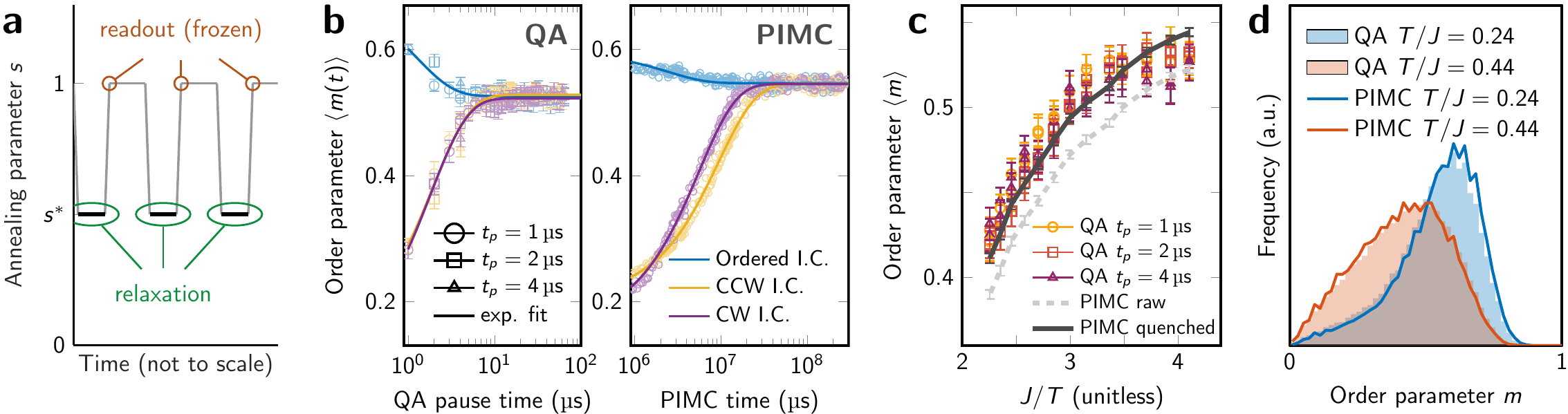}%
  \caption{{\bf Convergence of order parameter for 1440-spin lattice.} %
    {\bf a}, The QA protocol alternates between relaxation and readout, forming a chain of reverse annealing cycles, each containing a pause of length $t_p$ during which relaxation occurs in the simulated model, parameterized by $s^*$ (Methods). This breaks the simulation into discrete units, allowing observation of QA relaxation as in the Markov-chain PIMC simulation.
    {\bf b}, Starting from ordered, CCW, and CW initial conditions, QA and PIMC estimates of the time-dependent real order parameter $\braket{m(t)}$ converge to terminal values $\braket m$ for parameters $\Gamma/J = 0.736$ and $T/J = 0.244$ ($s=0.38$, $T = \SI{13.7}{mK}$).  Exponential fits capture the convergence near equilibrium, where statistical estimates become independent of initial condition.  %
    {\bf c}, QA estimates of $\braket m$ closely agree with quenched PIMC results (with local excitations removed) over a range of temperatures for $\Gamma/J = 0.736$, for two mappings of the lattice onto the QA processor.  We attribute the underestimate of $0.01$ at low temperatures to disorder in the Hamiltonian (Methods). %
    {\bf d},~Histograms of $m$ at high and low temperatures for $\Gamma/J = 0.736$ show accurate simulation of the entire distribution (PIMC data quenched). %
    All error bars are 95\% confidence interval on the mean.
  } %
  \label{fig:2}
\end{figure*}

A major challenge in probing QA dynamics is the lower bound on experimental timescales imposed by control circuitry\cite{Roennow2014,Albash2017b}, in this case $\SI{1}{\micro s}$.  To lengthen QA convergence timescales we take the novel approach of crafting initial states based on global topological obstruction in the pseudospin (Fig.~\ref{fig:1}d).  In addition to the {\em ordered} state, where the pseudospins are aligned, we initialize the system such that the pseudospin winds in a counterclockwise (CCW) or clockwise (CW) direction along the periodic dimension of a cylindrical boundary condition.  This can be thought of as a system-spanning vortex-antivortex pair whose centers are located beyond the two open boundaries.  The result is a ground state of the classical system; the corresponding quantum state is metastable, separated from the ground state by a spatial topological obstruction.  Escape from this state, which requires global unwinding and the introduction of a topological excitation within the lattice, provides a clear and relatively slow statistical signal of relaxation in a complex system.  Crucially, the twisted initial states make QA relaxation slow enough to observe experimentally in a rich, complex condensed matter system (this is in contrast to slowing QA with local cluster bottlenecks as in previous studies\cite{Denchev2016,Albash2017b}; in this case the clusters can be handled easily by a sufficiently sophisticated classical algorithm\cite{Mandra2016a,Albash2017b}).  The ordered and twisted states respectively maximize and minimize $m$ at $2/\sqrt 3$ and $0$ (Fig.~\ref{fig:1}d).  Our experimental objective is to compare relaxation from these initial conditions toward equilibrium in PIMC and QA.

\begin{figure*}
  \includegraphics[scale = .75]{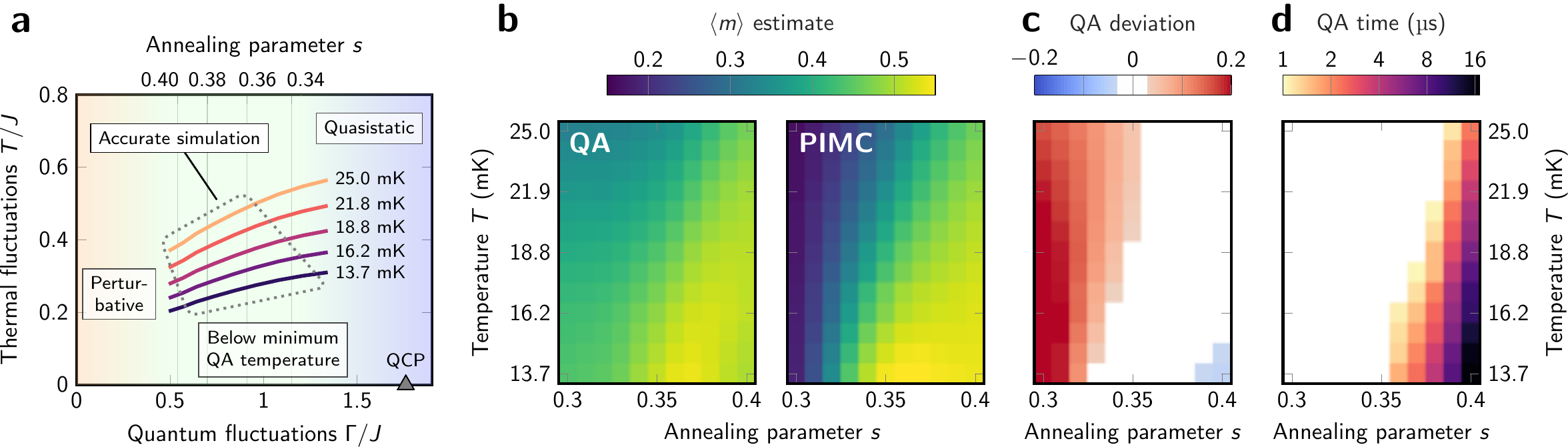}%
  \caption{{\bf Region of accurate simulation and resolvable dynamics.} %
    {\bf a},~Our QA experiments are parameterized by annealing parameter $s$ and physical temperature $T$; these map to the familiar $(T/J,\Gamma/J)$ plane as shown.
    {\bf b},~Estimates of $\braket m$ are shown for QA and PIMC quenched samples for a range of $T$ and $s$. %
    {\bf c}, QA deviation from PIMC estimates is shown.  When relaxation is extremely fast, order in the system is severely overestimated due to ordering during the readout quench.  The white region falls within a tolerance of $0.03$ of the ground truth, which corresponds to our ``accurate simulation'' region in {\bf a}.  At low $T$ and high $s$, transverse field is small and ordering is weak, and therefore easily suppressed by inhomogeneities in the processor. %
    {\bf d}, We cannot resolve convergence timescales faster than $\SI{1}{\micro s}$.  Our parameter range is therefore further restricted, excluding the fastest-converging models, i.e., those with high $T$ and low $s$, shown in white.
  }\label{fig:3}
\end{figure*}

In PIMC, simulation time is naturally quantized into Monte Carlo sweeps, providing a fine-grained time series of system evolution.  In QA we mimic this by using a ``quantum evolution Monte Carlo'' protocol\cite{King2018} that alternates between relaxing and reading out the system (Fig.~\ref{fig:2}a).  To simulate the Hamiltonian (\ref{eq:ham}) at annealing parameter $s^*$, we initialize the QA processor in a specified classical state (Fig.~\ref{fig:1}d) at $s=1$, where both quantum and thermal fluctuations are negligible, and dynamics are frozen.  We turn on dynamics by rapidly reducing $s$ to $s^*$ and waiting for a pause time $t_p$, which ranges from $1$ to $\SI{4}{\micro s}$.  We then quench the system, rapidly increasing $s$ back to $1$ and reading out a projected state in the computational basis.  Repeating this process many times allows us to measure QA relaxation as a time series with microsecond resolution.

The average order parameter $\braket m$ provides a simple representative statistic to estimate; in Methods we consider other observables such as two-point correlation and winding number, and find that the quantum simulation accurately quantifies a subtle equilibrium winding asymmetry.  Fig.~\ref{fig:2}b shows time-series convergence of $\braket{m(t)}$ to a converged average $\braket m$ in QA and PIMC from ordered, CCW, and CW initial conditions; each data point is the average of 600 experiments at $\Gamma/J=0.736$ and $T/J = 0.244$.  Quantum and classical simulations show qualitatively similar convergence for fixed $\Gamma$ and $T$, with the quantum simulation converging roughly three million times faster.  Operations such as programming and readout are excluded from QA time, as dynamics are effectively frozen; this is confirmed by supplemental experiments.  Across a range of temperatures, the slight but systematic QA overestimate of $\braket m$ (Fig.~\ref{fig:2}c) is explained by the annihilation of local excitations during the QA readout quench.
To account for the QA quench we apply a local quench to PIMC output; unquenching QA output provides a good match with unquenched PIMC samples (Methods).
QA faithfully simulates not only the average value of $m$, but the entire distribution (Fig.~\ref{fig:2}d).

We simulate the system over a range of QA parameters: annealing parameter $s$ ranges from $0.30$ to $0.40$, and physical temperature $T$ ranges from $\SI{13.7}{mK}$ to $\SI{25.0}{mK}$.  This maps to a region in the $(\Gamma/J,T/J)$ plane (Fig.~\ref{fig:3}a).  Fig.~\ref{fig:3}b--c show quenched QA and PIMC estimates of $\braket m$ and the deviation between them.  Across a large swath of the parameter space, QA and PIMC estimates agree to within a tolerance of $0.03$, indicating accurate simulation.  When $\Gamma$ and $T$ are high, dynamics are fast and QA severely overestimates order in the system due to global ordering during the $\SI{1}{\micro s}$ readout quench.  When $\Gamma$ is low, pseudospin interactions become weak, dropping off as a fourth-order perturbation in $\Gamma$.  Consequently order-by-disorder is easily suppressed by unintended disorder in the QA Hamiltonian, particularly at low temperature (Methods), leading to a deviation between QA and PIMC simulations.

\begin{figure}  \includegraphics[scale=.8]{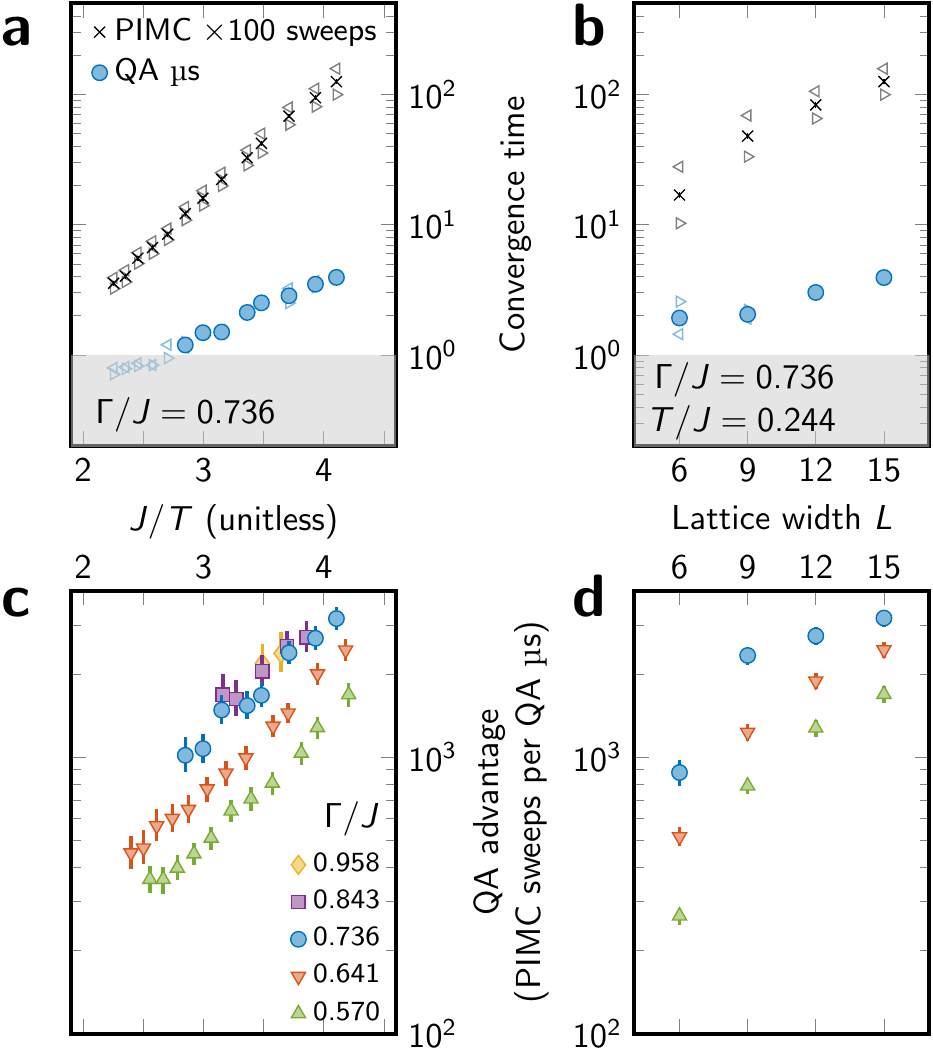}
  \caption{{\bf Scaling of convergence time and QA speedup.} %
    {\bf a--b},~Convergence time for both QA and PIMC as a function of inverse temperature $J/T$ ({\bf a}) and lattice width $L$ ({\bf b}).  Triangles $\triangleleft$ and $\triangleright$ indicate times for CCW and CW initial states respectively; other markers indicate geometric mean.  QA data is discarded if the estimate of $\braket m$ is not accurate to within $0.03$, or either CCW or CW convergence time is $< \SI{1}{\micro s}$ (shaded region).
    {\bf c--d},~QA advantage over PIMC, given as the ratio of convergence times, increases as $T$ decreases ({\bf c}), as quantum fluctuations increase ({\bf c}), and as system size increases ({\bf d}).  Temperatures shown in {\bf d} are minimum for which QA results are accurate ($J/T \approx 4.2$ in each case).  Scaling in $L$ ({\bf b}, {\bf d}) is given in terms of PIMC sweeps to show relaxation dynamics rather than computation time.  All filled data points have $95\%$ CI bootstrap error bars, often smaller than the marker.  At $\Gamma/J=0.736$, $T/J=0.244$, QA relaxation is three million times faster than PIMC on a CPU (Methods).
  }%
  \label{fig:4}
\end{figure}

To extract a convergence time for a chosen simulator and initial condition, we fit time-series data $\braket{m(t)}$ to a simple exponential process
\begin{equation}\braket{m(t)} = (m_0-m_f)e^{-t/\tau}+m_f
\end{equation}
with fitting parameters $m_0$, $m_f$, and $\tau$.  The natural timescale of interest is $\tau$, but for practical reasons we use cutoff time $t$ such that $m(t)-m_f=0.05$; this assures quality of data and does not affect scaling of convergence time (Methods).  It also allows us to discriminate between fast local ordering, which dictates $m_0$, and slower global unwinding in the lattice.
We consider only the longest timescales, which come from CCW and CW initial conditions, and use their geometric mean as a single representative convergence time (Fig.~\ref{fig:3}d).  Timescales less than the minimum experimental resolution of $\SI{1}{\micro s}$ cannot be measured reliably and are therefore discarded.  This is analogous to the requirement of well-resolved optimal anneal times for faithful scaling studies in QA optimization\cite{Roennow2014,Denchev2016,Albash2017b}.  We emphasize that as in other experimental investigations of QA performance scaling, we have ignored non-relaxation times such as programming and readout since they give the false appearance of flat scaling.

We now consider variation of QA and PIMC convergence time as a function of temperature $T$, transverse field $\Gamma$, and lattice width $L$.  We employ continuous-time PIMC code that updates four-qubit FM chains collectively, thereby obviating any local cluster bottleneck that might favor QA.  This and other implementations of PIMC are detailed in Methods.  Convergence timescales of both QA and PIMC increase as temperature decreases and as  lattice width $L$ increases (Fig.~\ref{fig:4}a--b).  We show PIMC convergence in Monte Carlo sweeps, viewing Metropolis dynamics as a quasi-physical process.
The dynamical advantage of QA, given as the ratio of PIMC and QA convergence times, varies systematically not only in $T$ and $L$, but also in $\Gamma$ (Fig.~\ref{fig:4}c--d).  Thus the computational advantage conferred by the quantum hardware increases as the simulations become harder (increasing $L$ and decreasing $T$) and as quantum mechanical effects increase (increasing $\Gamma$).  Note that while we have attributed a low-temperature suppression of QA order to Hamiltonian disorder, a similar effect---along with a distortion of scaling---could arise from a pernicious and systematic mismeasurement of $T$; this scenario is unlikely, as we have directly measured effective qubit temperature {\em in situ} and observe consistent slowing of QA convergence across the lower temperatures (Methods).

In the presence of noise and finite temperature we expect only finite-range quantum tunneling \cite{Denchev2016}.  Indeed, QA and PIMC scaling in $T$---governed by global unwinding of the lattice---resemble thermal activation $t \propto e^{a/T}$ with different exponents $a$ (Fig.~\ref{fig:4}a).  Given the four-qubit updates employed by PIMC, this indicates that the computational value of QA tunneling in this setting extends beyond the four-qubit chains used in collective PIMC updates, which is consistent with previous experiments \cite{Lanting2014,Denchev2016,Boixo2016}.

In contrast to previous benchmarking studies\cite{Denchev2016,Mandra2016a,Albash2017b}, our work shows not only a large absolute advantage, but also a scaling advantage over the corresponding cluster-aware classical method, on inputs that are of independent interest.  Escape from an obstruction is essential to the experiment---it lengthens QA timescales into the experimentally resolvable regime---and for this reason we have only compared against PIMC relaxation with a fixed Hamiltonian.  Many methods, ranging from trivial to highly sophisticated, can hasten escape in both PIMC and QA.  For example, the initial obstructed state can simply be randomized away, or nonlocal cluster moves based on dimer configurations in the classical ground state manifold can be interleaved with fixed-Hamiltonian relaxation.  These approaches would affect scaling in both PIMC and QA, and will require closer investigation when QA relaxation timescales become long.  We expect the development and analysis of such hybrid QA algorithms to be an active area of research in the coming years \cite{Preskill2018,Morley2019}.  Conversely, the experimental QA protocol can also be applied to PIMC simulations: while quenching according to QA experimental parameters roughly maintains---in both QA and PIMC---dynamics and escape timescales, fast and frequent PIMC quenches lead to a highly non-equilibrium dynamics and can accelerate escape (Methods).

In conclusion, we have experimentally demonstrated a computational scaling advantage in simulating frustrated magnetism with a programmable superconducting quantum annealing processor, as well as a large absolute advantage in convergence time.  To our knowledge, this is the first experimental evidence of a scaling advantage for QA over PIMC in a 2-local stoquastic Hamiltonian, and the first measurement of non-perturbative QA relaxation in a large frustrated system.  Far from being an artificial benchmark, the simulated lattice demonstrates the exotic topological phenomena that can arise in frustrated quantum Ising systems.  As Monte Carlo inference is an effective tool in the study of both idealized frustrated systems\cite{Moessner2001,Isakov2003} and real frustrated magnetic compounds such as Ca$_3$Co$_2$O$_6$~[\onlinecite{Kamiya2012}] and TmMgGaO$_4$ [\onlinecite{Li2019}], our experiment is closely related to real-world applications.

These results constitute an encouraging milestone: a programmable quantum system can simulate quantum condensed matter far faster than the corresponding classical method, and with better scaling in problem size and hardness.  Extensions of this work abound: related phenomena of great interest include material properties in the vicinity of a quantum critical point\cite{Sachdev2011}, and dynamics of monopole excitations in artificial spin ice \cite{Farhan2019}.  Simulating these near the zero-temperature limit in QA would benefit from processors with more flexible lattice connectivity, higher coupling energy, lower noise, and improved projective readout.  These programmable quantum simulations may ultimately be applied to the design of exotic new materials that are just beyond the computational horizon.  The advantage reported in this work also opens the door to hybrid approaches that could be used to accelerate high-performance computing tasks.  As various quantum computing technologies mature, we anticipate similar scaling advantages in the simulation of quantum systems---such demonstrations are crucial waypoints for the field as a whole.

\section*{Acknowledgments}

We are grateful to Tameem Albash,  Itay Hen, Daniel Lidar, Vadim Smelyanskiy and Karsten Kreis for helpful conversations.

\bibliography{paper_fftfim}%
\clearpage

\onecolumngrid
\appendix


\tableofcontents

\subsection{Fully-frustrated Ising model and initial conditions}

The magnetic system studied herein is a square-octagonal lattice whose spins are partitioned into four-spin ferromagnetically coupled chains (Fig.~\ref{fig:1}a).  Each chain is contained in an octagon; the remainder of the couplers are antiferromagnetic.  Ferromagnetic and antiferromagnetic couplings have respective strengths of $J_{ij}=-1.8$ and $J_{ij}=1$.  Each plaquette of the lattice has three antiferromagnetic bonds, leading to a {\em fully-frustrated} Ising model.

Like the closely-related triangular antiferromagnet, this lattice has extensive ground state entropy in the classical case, and no long-range order in the zero-temperature limit.  Upon addition of a perturbative transverse field $\Gamma$, the system undergoes order-by-disorder; the resulting quantum system has, to first order perturbation, sixfold ground state degeneracy.  As a result, in the limit $T\approx 0$, $\Gamma\approx 0$, the system has a mapping to a six-state XY clock model:  In this dual {\em pseudospin} lattice, each plaquette of the triangular lattice is assigned a complex pseudospin
\begin{equation}\label{eq:pseudospin}
\psi_j = \sigma_{j1}^z + \sigma_{j2}^ze^{2\pi i /3} + \sigma_{j3}^ze^{4\pi i /3}.
\end{equation}
The coupling between these pseudospins is described in the Methods section of previous work\cite{King2018}.  This plaquette pseudospin can be extended to the square-octagonal system by replacing the $\sigma^z$ operator on a spin by the magnetization of a four-qubit chain.  In the square-octagonal case, the pseudospin coupling is driven by the energetic contributions of four-qubit GHZ states $(\ket{\uparrow\uparrow\uparrow\uparrow}+\ket{\downarrow\downarrow\downarrow\downarrow})/\sqrt{2}$ on the chains, rather than single-qubit superposition as in the triangular AFM.

  In the perturbative limit the quantum system is accurately described to first order by an effective six-state 2D XY model.  Consequently we see an ordered phase at low temperature and an extended critical phase, bounded by topological phase transitions of 2D XY type\cite{Jose1977,Jalabert1991,Moessner2001}.  As $\Gamma/J$ is increased the 2D XY approximation fails, and the finite-temperature topological phase transitions give way to a quantum phase transition in the 3D XY universality class; at finite system size a crossover is observed \cite{Moessner2001,Isakov2003,King2018}.

The spins of the triangular AFM can be divided into three sublattices such that no two spins in the same sublattice are coupled.  Each perturbative ground state of the TFIM---in both the triangular and the square-octagonal case---orders the sublattices as $\ket{\uparrow\downarrow\updownarrow}$ in some permutation.  In contrast, we use an  $\ket{\uparrow\uparrow\downarrow}$ ordering for our ordered initial condition.  This gives a classical analog of the $\ket{\uparrow\downarrow\updownarrow}$ states: the pseudospin field $\psi_j$ is constant and large in magnitude, and maximizes---for classical states---the number of ``floppy'' or ``flippable'' spins or chains that experience no net effective field.

There are six such ordered initial states, with sublattice orderings $\uparrow\uparrow\downarrow$ or $\uparrow\downarrow\downarrow$ in some order; these states have complex order parameter $\psi$ being $e^{ik\pi/3}\cdot 2/\sqrt 3 $, $k=1,\ldots, 6$.  Accordingly, we denote the states by $S_k$, $k=1,\ldots, 6$.  To construct the CCW initial condition for a lattice on $6m$ rows, for $k=1,\ldots, 6$ we assign the spins of rows $1+(k-1)m$ to $km$ the value they are given in $S_k$.  In this way we construct a state in which the pseudospin winds in a full rotation along the periodic dimension of the cylinder.

At finite temperature, chain-break excitations play an important role in the square-octagonal lattice.  In the cylindrical system, the clockwise and counterclockwise directions become asymmetric, as can be seen in the difference between relaxation from CCW and CW initial conditions (Fig.~\ref{fig:2}).

We study lattices on $4L(2L-6)$ spins with cylindrical boundary condition, with lattice width $L$ chains from one open boundary to the other, and $2L-6$ chains around the periodic dimension.  We study instances with $L\in\{6,9,12,15\}$, the largest system having 1440 spins.  The presence of inoperable qubits prevents us from repeating the study of an 1800-spin system\cite{King2018}.

\begin{figure}
  \includegraphics[scale=.75]{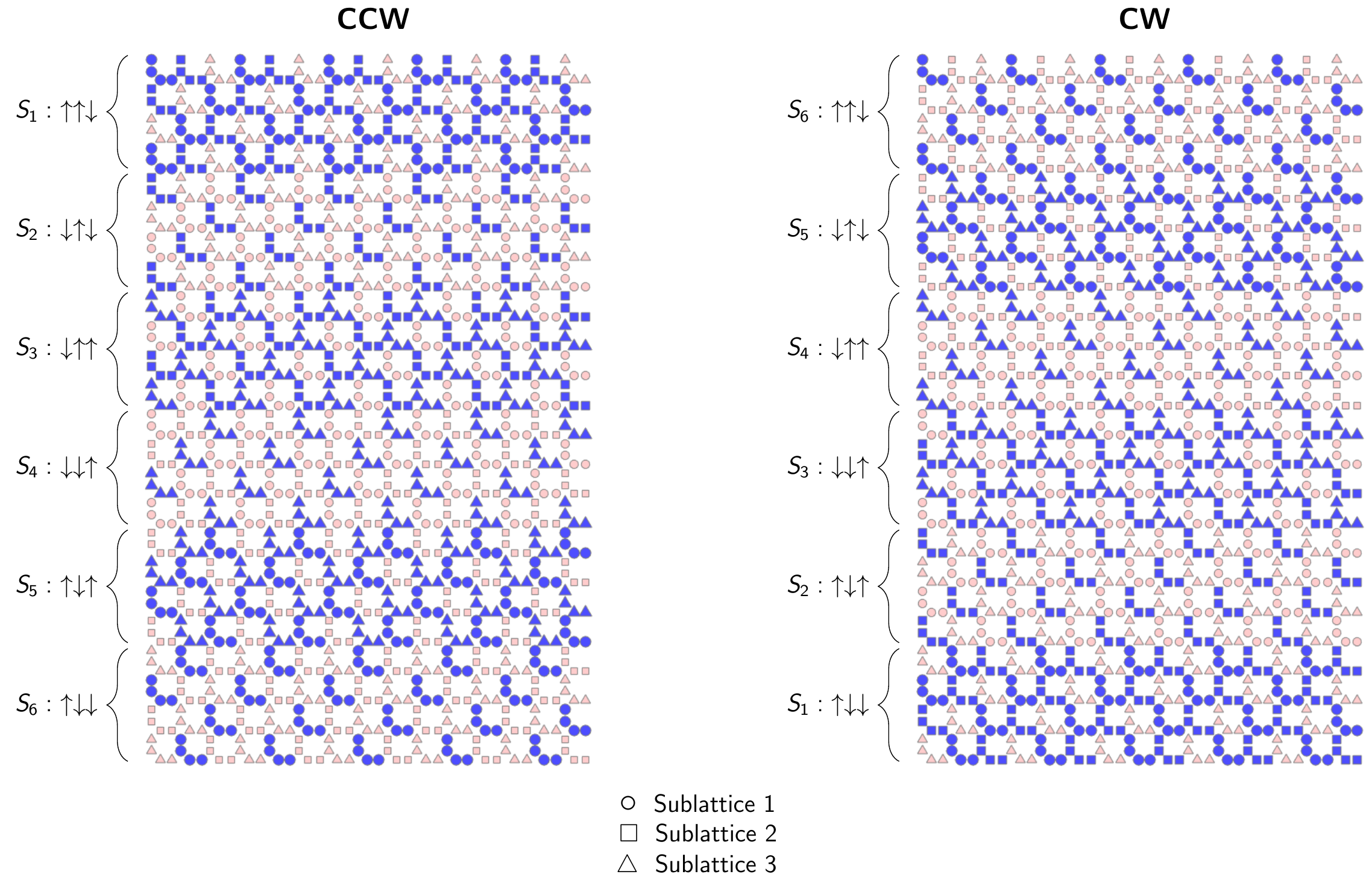}
  \caption{{\bf CCW and CW twisted initial conditions.}  Shown are examples of twisted initial conditions for the largest system size studied: a $24\times 15$ lattice of four-qubit chains.  The lattice is divided into six stripes of four rows each; each stripe is assigned to a different ordered initial condition $S_k$ winding around in a complete rotation.  The three sublattices, derived from the triangular antiferromagnet, are shown with different marks: circles, squares, and triangles respectively.
    The state $S_k$ has complex order parameter $e^{ik\pi/3}\cdot 2/\sqrt 3 $.  The resulting CCW and CW states are ground states of the classical potential.
    The winding of the pseudospin provides a topological obstruction from which QA and PIMC must escape.}\label{fig:initialconditions}
\end{figure}

\subsubsection{Winding number}

We quantify topological winding of the pseudospin field as follows.  We take a pseudospin field $\mathcal P$ as a square lattice whose rows alternate between neighboring octagonal and square plaquettes; a cylindrical lattice with $L=15$ made up of a $15$-by-$24$ lattice of chains then becomes a $28$-by-$24$ square grid of plaquette pseudospins.  We then take the two-dimensional Fourier transform
\begin{equation}
  M(a,b) = \sum_{j=0}^{L_1-1}\sum_{k=0}^{L_2-1}\exp(2\pi i\tfrac{j a}{m})\exp(2\pi i\tfrac{kb}{n})\mathcal P_{j,k}
\end{equation}
where $L_1$ and $L_2$ are the dimensions of the square lattice representation of $\mathcal P$.  Note that in absence of boundary conditions, $M(0,0)/(L_1L_2)$ is the complex order parameter $\psi$.  For an integer winding number $w$, we define
$$f(w) = \sqrt{ \sum_{a=0}^{L_1-1}\left(\frac{|M(a,w)|}{L_1L_2} \right)^2}.$$
This gives us 24 possible values of $w$.  The ordered, CCW, and CW states have peaks in $f(w)$ at $w=0$, $w=1$, and $w=-1$.

\subsection{Quantum annealing processor}\label{sec:qpu}

Quantum annealing experiments were performed on a quantum processing unit with a similar circuit architecture to the processor used in previous quantum simulation experiments\cite{King2018,Harris2018,Bunyk2014} but using a lower-noise fabrication process.  In the previous system\cite{King2018,Harris2018}, all 2048 radio-frequency SQUID flux qubits and 6016 couplers were operational.  In this experiment, 2030 qubits and 5909 couplers were operational; consequently we studied systems of up to 1440 spins instead of up to 1800 spins.

Each lattice size is embedded in two different ways in the qubit connectivity graph of the QA processor, differing primarily by a half rotation.  As in previous experiments\cite{King2018}, we maintain a calibration refinement to compensate for static crosstalks and minimize time-dependent drift in the Hamiltonian.  This refinement minimizes variance in individual qubit magnetizations using flux-bias offsets.  Using fine-grained modifications of coupling energies, it minimizes variance among the statistics of isomorphic antiferromagnetic couplers that relate to one another either by rotation of the cylinder or by isomorphic location in the two embeddings.

Quantum annealing itself typically involves sweeping the annealing parameter $s$ from $0$ to $1$ in the Hamiltonian (\ref{eq:ham}), using a schedule of $J(s)$ and $\Gamma(s)$ such that $J(0)\ll \Gamma(0)$ and $J(1)\gg \Gamma(1)$ [\onlinecite{Kadowaki1998}].  This approach is closely related to adiabatic quantum computing\cite{Farhi2001,Albash2016a} performed at finite temperature.  In this work we use a QA processor to probe Hamiltonian (\ref{eq:ham}) at an intermediate value of $s$, rather than annealing $s$ gradually.
Experiments were run in a quantum evolution Monte Carlo loop\cite{King2018}, which mimics Markov-chain Monte Carlo but replaces the Markov chain update step with a reverse anneal.
The QPU is initialized with a classical state, which for the first reverse anneal is a specified input, and for later anneals is the output of the previous step. 
Each anneal, aimed at evolution of the Hamiltonian $H(s^*)$, begins with $s=1$, reduces $s$ to $s^*$ over $1-s$ microseconds, pauses for time $t_p$, then increases $s$ from $s^*$ to $1$ over $1-s$ microseconds (detail in Fig.~\ref{fig:quenchsensitivity}).
The major benefit of using QEMC in this experiment is that it allows fine-grained examination of intermediate points of the relaxation from initial condition to the converged distribution.
At the end of each anneal, we insert a pause of $\SI{10}{ms}$ to minimize potential heating and sample-to-sample correlation.  These pauses dominate overall experimental time and we have not attempted to optimize them as a parameter.  
Experiments were run for a range of temperatures, annealing parameters, and lattice sizes.  Estimates of dynamics, with the exception of Fig.~\ref{fig:3}d for $s\leq 0.35$, were drawn from $300$ QEMC chains of $16$ samples each, for each of two lattice embeddings.  Equilibrium estimates, and data for $s\leq 0.35$ in Fig.~\ref{fig:3}d, were drawn from $60$ chains of $128$ samples; for equilibrium estimates we discard the first half of the chain as burn-in.

\subsubsection{Extracted QA timescales}\label{sec:cutoff}

Since measured QA timescales are often near the minimum experiment time of $\SI{1}{\micro s}$, we must determine a reasonable cutoff for measurements in which we have confidence.  This goes hand in hand with our definition of convergence time.  We fit observations to a simple exponential process
\begin{equation}m(t) = (m_0-m_f)e^{-t/\tau}+m_f
\end{equation}
where $m_0$, $m_f$, and $\tau$ are fitting parameters.  The long-time limit $m_f$ is well defined as the equilibrium estimate of $\braket m$.  However, the role of $m_0$ is not as clear.  Although the fit function agrees very well with the observed data, the convergence process is not a simple exponential.  Rather, there are at least two significant timescales.  This is explained by the fact that the CCW and CW initial conditions have many four-qubit chains experiencing zero net field, meaning that they can flip freely.  As such, $m$ quickly relaxes to a nonzero distribution.  On the much longer timescale of interest, the system escapes the topological obstruction to approach the equilibrium value of $\braket m$.  The faster timescale is not seen in QA data, meaning that $m_0$ is not well defined.
Therefore we define our convergence time to be the time at which the fit function converges to within a cutoff value of its equilibrium value.

\begin{figure}
  \includegraphics[scale=.8]{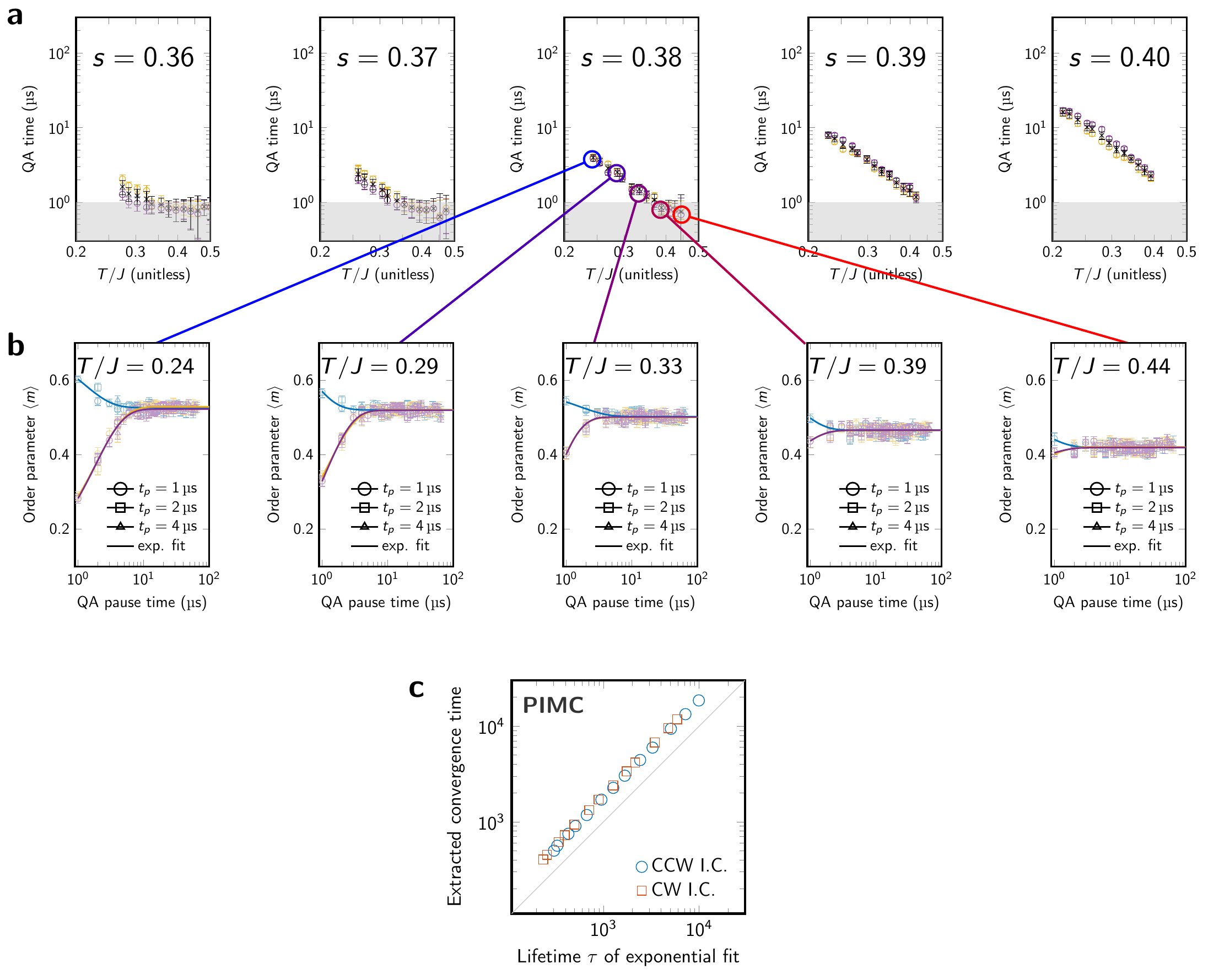}
  \caption{{\bf Resolvable and unresolvable convergence times.}  We consider any convergence times, given as the time at which the exponential fit function is $0.05$ away from its terminal value, to be resolved only if it is greater than $\SI{1}{\micro s}$.  This condition is only relevant for QA, since all PIMC times are resolvable.  {\bf a}, QA times are extracted from exponential fits for a variety of annealing parameters $s$ and temperatures $T/J$.  Yellow and purple marks are from CCW and CW initial conditions, respectively; black marks are the geometric mean of the two, which we use as a single representative timescale.  We discard any data for which either CCW or CW convergence time is less than $\SI{1}{\micro s}$. 
    {\bf b},~Timescales are extracted from exponential fits to data from CCW and CW initial conditions.  In the middle panel, for each of CCW and CW only three data points are significantly separated from the terminal value: $1\times \SI{1}{\micro s}$, $1\times \SI{2}{\micro s}$, and $2\times \SI{1}{\micro s}$ ($1\times \SI{2}{\micro s}$ and $2\times \SI{1}{\micro s}$ match closely, indicating that our measurement of relaxation time is accurate).  These data are supported by the fact that convergence varies smoothly as a function of temperature, but for higher temperatures even the first data point is indistinguishable from sampling error, thus we cannot resolve a credible convergence timescale.
  {\bf c},~Our extracted timescales, given by time to reach a cutoff of $0.05$ in the exponential fit, is motivated by assuring statistical significance of the fit itself for small QA times.  Here we scatter PIMC convergence times, extracted with the same methods, against lifetime $\tau$ of the exponential fit, showing strong agreement for $s=0.38$ over a variety of temperatures.}\label{fig:timescales}
\end{figure}

The choice of this cutoff is motivated by the quality of data, as illustrated in Fig.~\ref{fig:timescales}.  The chosen cutoff of $0.05$ is significantly larger than variation of individual QA estimates of $m(t)$.  This means that any measured QA convergence time of greater than $\SI{1}{\micro s}$ is supported by strong evidence.  Reducing the cutoff could lead to the inclusion of extracted cutoff times with weak supporting evidence.

\subsubsection{Insensitivity to quench rate}\label{sec:quench}

\begin{figure}
  \scalebox{0.87}{%
  \begin{tikzpicture}\sf
    \node at (-6,0.2) {\includegraphics[scale=0.8]{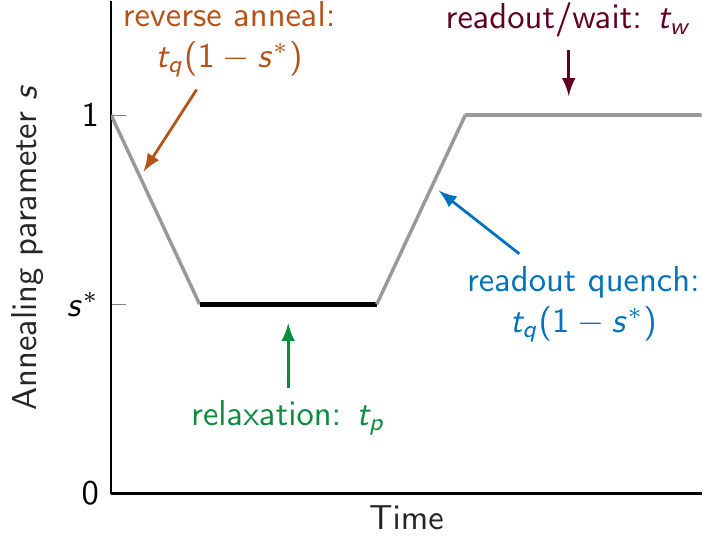}};
    \node at (0,0) {\includegraphics[scale=0.8]{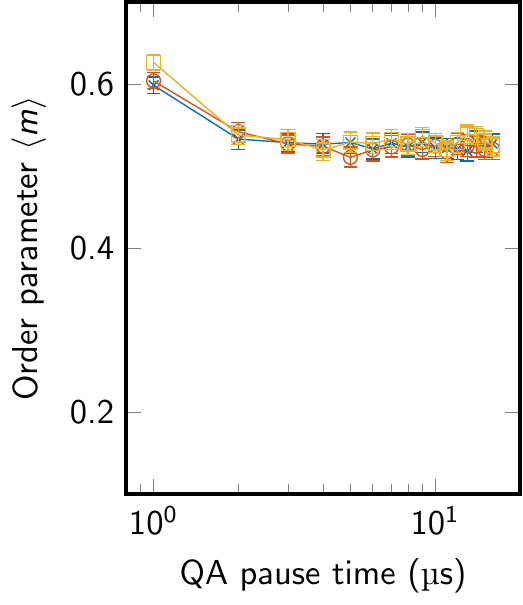}};
     \node at (4.7,0) {\includegraphics[scale=0.8]{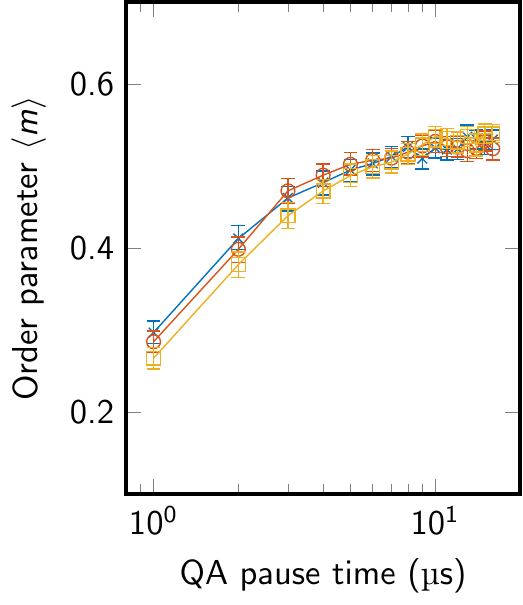}};
     \node at (9.4,0) {\includegraphics[scale=0.8]{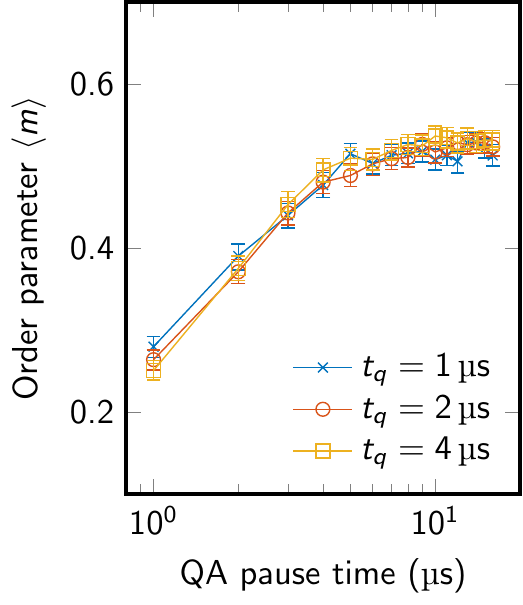}};
     \node[anchor = east] at (2,2) {\bfseries ORDERED};
     \node[anchor = east] at (6.7,2) {\bfseries CCW};
     \node[anchor = east] at (11.4,2) {\bfseries CW};     
   \end{tikzpicture}
   }
   \caption{{\bf Insensitivity to quench rate.} To support the claim that relaxation during quench and reverse anneal phases is insignificant, and therefore this time should be disregarded in analysis, we double and quadruple the amount of time spent during these operations.  A systematic trend of ``faster'' relaxation for longer quench time would suggest that out-of-pause relaxation is significant; we observe no such effect.  Data shown are for $t_p= \SI{1}{\micro s}$, $s=0.38$, $T = \SI{13.7}{mK}$ as in Fig. \ref{fig:2}.  Time spent during each of the reverse anneal and quench phases is $t_q(1-s)$, in this case $\SI{0.62}{\micro s}$, $\SI{1.24}{\micro s}$, $\SI{2.48}{\micro s}$.}\label{fig:quenchsensitivity}
\end{figure}
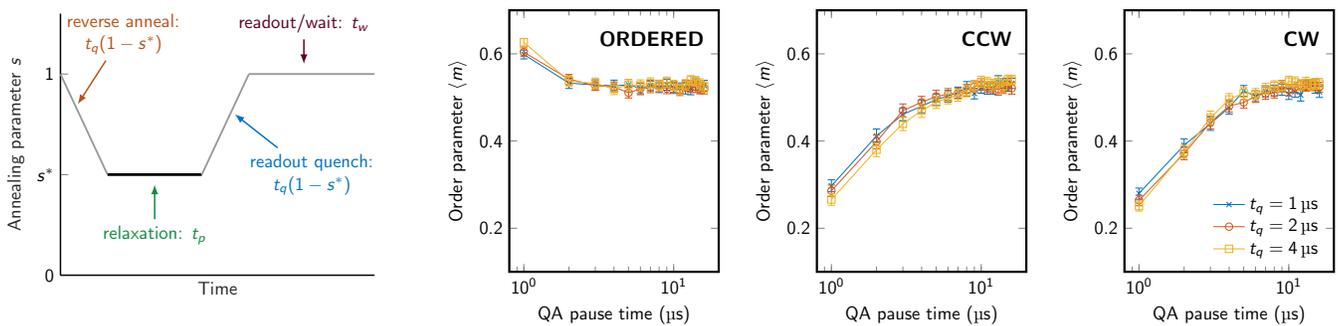

In this work we have measured QA relaxation time using only the relaxation pause portion of the QEMC protocol (Fig.~\ref{fig:quenchsensitivity}).  To justify this approach, we provide evidence that relaxation outside the pause is negligible.  We do so by showing that QA relaxation is insensitive to changes in the quench rate.

Since we observe that QA convergence time varies exponentially in $s$ over the observable range of roughly $0.36 \leq s \leq 0.40$, we consider the following oversimplified but illustrative model of relaxation far from equilibrium.  Let us momentarily consider relaxation as an $s$-dependent task that is performed at a rate proportional to $e^{-\alpha s}$, for $\alpha$ chosen such that the system relaxes twice as fast at $s=0.38$ as at $s=0.39$.  Since the quench and reverse anneal portions of the QEMC protocol have $|ds/dt| = 1/(\SI{1}{\micro s})$ in our experiments, and pause time $t_p \geq \SI{1}{\micro s}$, the total change in $m$ during each of the reverse anneal and quench phases is less than $2\%$ of the change in $m$ during the pause.

This model is vastly oversimplified but roughly in line with our observations.  To show experimentally that out-of-pause relaxation is negligible in our experiments, we perform a spot-check at $s=0.38$, $T=\SI{13.7}{mK}$ with pause time $t_p=\SI{1}{\micro s}$.  We use reverse anneal and quench rates of $|ds/dt| = 1/(\SI{1}{\micro s})$, $1/(\SI{2}{\micro s})$, and $1/(\SI{4}{\micro s})$ (Fig.~\ref{fig:quenchsensitivity}).  The consistency of order parameter convergence as a function of pause time indicates that out-of-pause relaxation is negligible.

\subsection{Measurement of QA parameters}

The comparison of QA and PIMC relaxation requires detailed measurement of the QA processor parameters, and mapping to the resulting parameters of the implemented TFIM. In particular, we require measurements of qubit body inductance $L_q$, capacitance $C_q$, critical current $I_c$, and the two terms in the quantum annealing Hamiltonian: qubit tunneling energy $\Gamma_q(s)$, and qubit coupling energy $J_q(s)$, for the range of $s$ we studied.

\subsubsection{Qubit model parameters}

We measure $L_q$ and $I_c$ by measuring qubit persistent current as a function of $s$ in the the regime $\Gamma_q \ll J$ and fit to a classical model of the radio-frequency (rf) SQUID~\cite{Harris2010a}. We measure $L_q = 284$ pH and $I_c = 2.34\ \mu{\rm A}$.

We then measure $\Gamma_q(s)$ and device capacitance with qubit tunneling spectroscopy. For a given value of $s$, we perform single qubit spectroscopy with a single attached probe qubit as described in Ref.~\onlinecite{Berkley2013}. At this value of $s$, we sweep the qubit through degeneracy and fit the energy eigenspectrum to a one-dimensional dispersion relation. This gives a measurement of $\Gamma_q(s)$ and qubit persistent current, $I_p^q(s)$. We perform these measurements on 500 qubits for a range of $s$ and fit the resulting data to a SQUID model to obtain a best fit capacitance of $C_q = \SI{113.7}{\femto F}$. Figure~\ref{fig:params}a shows these measurements along with the best fit to a SQUID model.

Finally, we calibrate $J_q(s)$ by performing qubit tunneling spectroscopy on coupled two-qubit systems. When two qubits are coupled with a term $J_q(s)\sigma_1^z\sigma_2^z$, then $E_2(s) - E_1(s) \equiv 2J_q(s)$ where $E_1(s)$ and $E_2(s)$ are the first and second excited state eigenenergies of the two-qubit system, respectively, at $s$. Note that this holds for all values of $s$, even when $\Gamma_q \approx J_q$. To perform spectroscopy, we choose a value of $s$ and attach a third probe qubit to one of the two coupled qubits.  We then measure the energy eigenspectrum of the system at degeneracy as described in Ref.~\onlinecite{Lanting2014}.  We measure 32 two-qubit pairs at a range of $s$ to estimate $J_q(s)$ versus $s$. Figure~\ref{fig:params}b shows these measurements. We also show the prediction for $J_q(s)$ from the SQUID model. We emphasize that the dashed line in Figure~\ref{fig:params}b is not a fit: we use the calibrated device parameters and obtain excellent agreement between the two-qubit spectroscopy and the predictions from the SQUID model.

The square-octagonal lattice studied in this work contains sets of four qubits connected with $J_{ij} = -1.8$. The standard calibration we describe above is typically done with qubits coupled with $J_{ij} \approx 1$. Changing $J_{ij}$ on couplers attached to a particular qubit causes an inductive loading shift which we compensate with an inductance tuner attached to the qubit~\cite{Harris2010a}.  We identified a small but systematic offset in the inductance compensation when tuning couplers to the strong ferromagnetic regime used in this study.  We quantify this offset by repeating the calibration of $L_q$ as described above, but with a coupler attached to the qubit tuned to $J_{ij} = -1.8$ rather than the typical $J_{ij}\approx 1$ we use. For the embedding used in this study, we measure an average change in qubit inductance of $\sim 1\%$. Specifically, the average qubit inductance increases by $2.9 \pm 0.29$ pH.  Taking two standard deviations in this measurement gives a 95\% confidence interval on the qubit body inductance of $L_q = 286.9\pm 0.58$ pH; this uncertainty is a leading source of error in the QA schedule.

\subsubsection{Qubit temperature}

We measure effective qubit temperature using a susceptibility measurement.  This qubit temperature systematically deviates from the nominal cryostat set point by roughly one millikelvin.  Fig. \ref{fig:params}c shows the relationship between nominal cryostat temperature and measured qubit temperature.  Per-qubit variation in effective temperature is approximately Gaussian with a standard deviation of roughly $\SI{0.5}{mK}$.  Error in the mean qubit temperature of roughly 2\% is estimated by taking the minimum and maximum over four independent temperature measurements.

\subsubsection{Background susceptibility and compensation}

\begin{figure}\includegraphics[scale=.8]{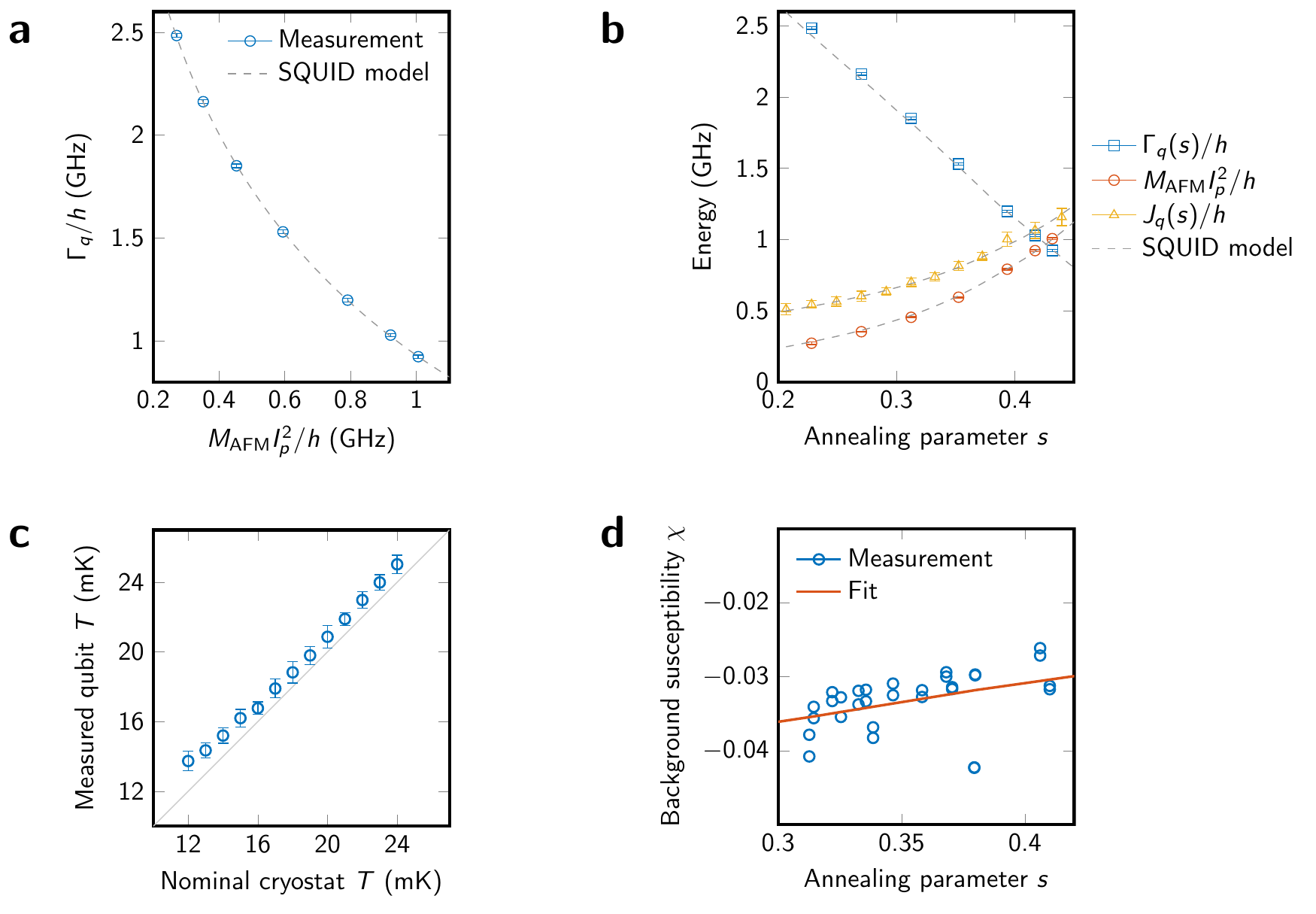}

  \caption{{\bf Measurement of QA parameters.} %
    {\bf a},  Measurements of $\Gamma(s)$ versus $M_{\rm AFM}I_p^2$. The solid lines show the best fit SQUID model.
    {\bf b},  Measurements of $\Gamma(s)$, $M_{\rm AFM}I_p^2$, and $J(s)$ versus $s$. The solid and dashed lines show the SQUID model.
    {\bf c},  QA experiments are peformed using nominal cryostat temperatures of between $\SI{12}{mK}$ and $\SI{24}{mK}$.  At each nominal temperature, an effective qubit temperature is measured via qubit susceptibility measurements.  Error bars indicate uncertainty in mean qubit temperature, taken from the minimum and maximum of four independent measurements at each nominal temperature.
    {\bf d}, Background susceptibility $\chi$ is measured through the study of two-qubit phase diagrams.  We use linear regression values for compensation across experimental values of annealing parameter $s$.
}\label{fig:params}
\end{figure}

A leading systematic deviation between our flux qubits and ideal spin-$1/2$ Ising moments is {\em background susceptibility} $\chi$, through which qubits mediate a next-nearest-neighbor coupling.  We estimate $\chi$ at a particular annealing parameter $s$ by measuring deviation from ideal in the phase diagram of two coupled qubits under independent fields $h_1$ and $h_2$ (Fig.~\ref{fig:params}d).  As in previous experiments (see \cite{King2018} Methods), we compensate for this behavior by tuning the programmed QA input couplings $J_{ij}$ such that upon application of background susceptibility, the effective Hamiltonian approximates the desired square-octagonal TFIM.  More formally, we denote the application of a background susceptibility $\chi$ to a classical Ising Hamiltonian $H$ by $f_\chi(H)$.  Using an iterative method, we find a Hamiltonian $H_{(-\chi)}$ such that $f_\chi(H_{(-\chi)}) \approx H$, as described in the Methods of previous work\cite{King2018}.  The relative tuning between AFM couplers is up to 2\%, and the relative tuning between FM couplers is up to 9\%.

\subsubsection{Effective spin-1/2 Hamiltonian}

\begin{figure}\includegraphics[scale=0.8]{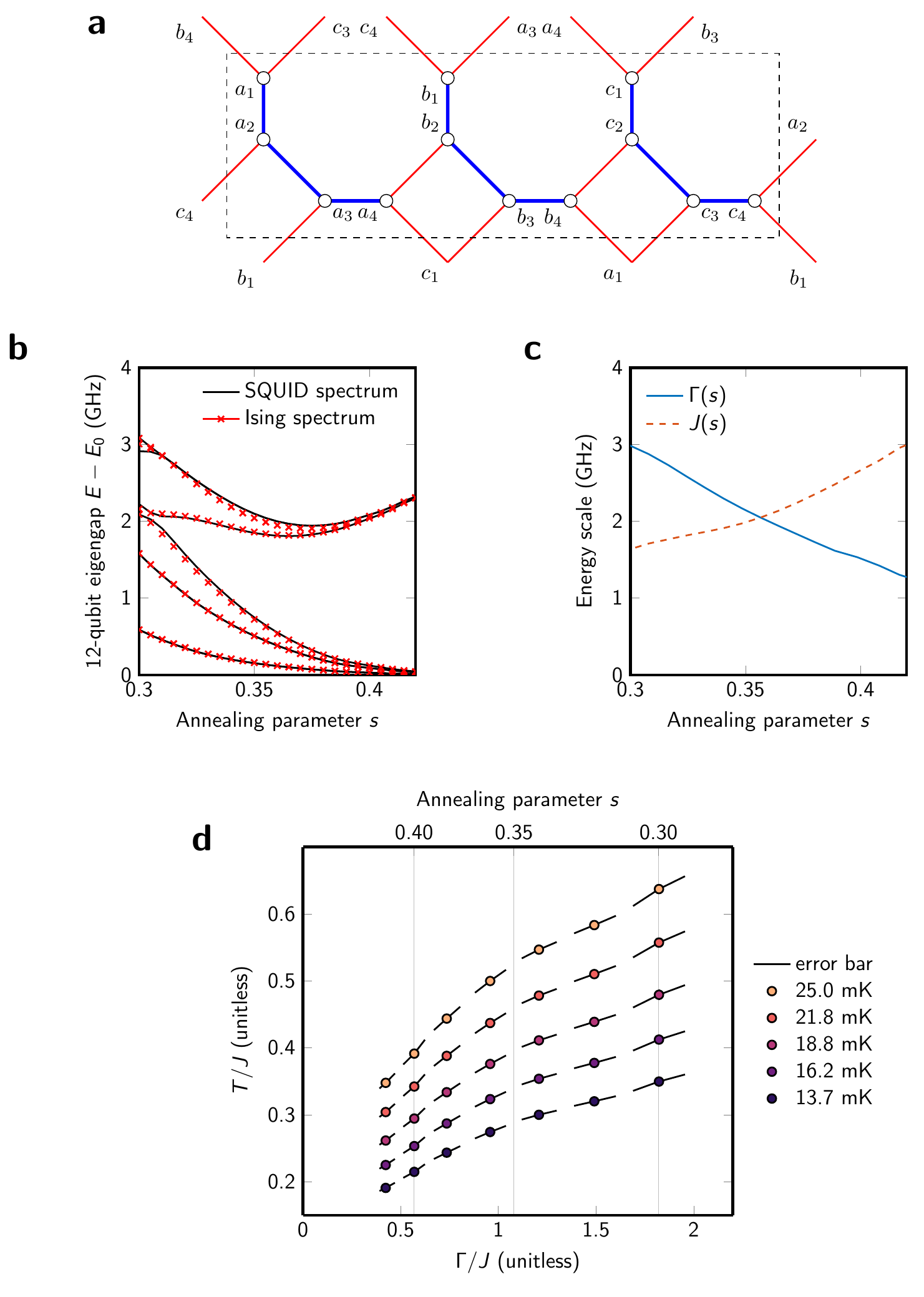}

  \caption{{\bf Effective transverse field Ising model in a network of SQUIDs.}
    {\bf a}, We study the spectrum of a minimal representative system: a 12-qubit square-octagonal lattice with periodic boundary.  The QA processor uses rf-SQUID flux qubits to implement an effective transverse field Ising model.
    {\bf b}, The lowest 9 energy levels of the SQUID spectrum are shown (some are degenerate), along with the spectrum of an effective Ising model whose parameters $\Gamma(s)$ and $J(s)$ are determined by fitting a weighted average of SQUID eigengaps.
    {\bf c}, The parameters $\Gamma(s)$ and $J(s)$ provide the annealing schedule of the effective Ising model as implemented in flux qubits.
  {\bf d}, Extracted Ising schedule is shown in the $(T/J,\Gamma/J)$ plane for a range of operating temperatures.  The two major uncertainties are in the average qubit body inductance and the average qubit temperature, which respectively give diagonal and vertical error bars (vertical bars are hidden by the markers).}\label{fig:squids}
  
\end{figure}

The QA processor approximately implements two-level spin-$1/2$ qubits in the TFIM using rf-SQUID flux qubits, which have more than two energy levels.  We have described measurement of the SQUID parameters.  To determine and validate an effective spin-1/2 Hamiltonian for the fully-frustrated square-octagonal lattice, we diagonalize the SQUID Hamiltonian for a 12-qubit square-octagonal system (three chains) with periodic boundaries (Fig.~\ref{fig:squids}a).  This gadget is used because it captures the three-sublattice ordering of the square-octagonal lattice, and is small enough to diagonalize easily.  It also captures local signatures of entanglement in agreement with measurements over a larger lattice (Sec.~\ref{sec:entanglement}).  Since background susceptibility is included in the SQUID model, we diagonalize $H_{(-\chi)}$ instead of $H$.  We perform approximate diagonalization using six energy levels per SQUID and retain 16 energy levels per four-SQUID chain.  We find best-fit values of $\Gamma(s)$ and $J(s)$ so that the first eight eigengaps of the Ising Hamiltonian approximately match the first eight eigengaps of the SQUID Hamiltonian (Fig.~\ref{fig:squids}b).  The resulting parameters admit a mapping from the QA annealing parameter $s$ to Ising parameters $\Gamma(s)$ and $J(s)$ (Fig.~\ref{fig:squids}c) with strong agreement between the low-energy spacing of the TFIM spectrum and the SQUID spectrum.

The main uncertainties in the QA schedule come from temperature $T$ and qubit body inductance $L_q$.  Fig.~\ref{fig:squids}d shows the QA schedule for several temperatures in the $(\Gamma/J, T/J)$ plane.  The uncertainty in average qubit temperature gives a vertical error bar (hidden by markers) and the uncertainty in average qubit body inductance ($\pm \SI{0.058}{\pico H}$) gives a diagonal error bar corresponding approximately to an uncertainty in $s$ of $\pm 0.007$.

\subsection{Path-integral Monte Carlo methods}

In this work we study several variants of path-integral Monte Carlo, a standard tool for estimating equilibrium statistics of systems in the transverse field Ising model.  To estimate statistics accurately PIMC acts in the limit of continuous imaginary time (CT-PIMC), and we focus on this continuous-time form in the main body of the paper.  While general-purpose PIMC code typically updates one spin at a time, collective tunneling of four-qubit chains is essential to the behavior of the square-octagonal lattice under study.   It is therefore natural that PIMC simulation can be accelerated by collectively updating four-qubit ferromagnetically-coupled chains.  We do so using Swendsen-Wang updates over four-qubit chains.

As with QA experiments, PIMC estimates are drawn from 600 independent repetitions for each parameter set.  Raw output is projected to the classical space by taking the top Trotter slice.  We perform up to $2^{22}$ PIMC4q sweeps to estimate equilibrium properties and convergence.

We also examined {\em stochastic series expansion} (SSE)---a Monte Carlo framework distinct from PIMC---specifically a cluster SSE algorithm that purports high speeds over frustrated lattices~\cite{Sandvik2003,Biswas2016}.  Our implementations proved uncompetitive by orders of magnitude compared to PIMC in the phases explored.

\subsubsection{Discrete-time PIMC and Trotter error}

The continuous-time PIMC approximation is based upon the Suzuki-Trotter transformation~\cite{Suzuki1976}
\begin{equation}
        Z = \mathrm{Tr}\left[\exp\left(-\frac{\mathcal{J}}{T}{\hat H}_P + \frac{\Gamma}{T} \sum_i \sigma^x_i \right)\right] = \mathrm{Tr} \left[\exp\left(-\frac{\mathcal{J}}{T M}{\hat H}_P\right)\exp\left(\frac{\Gamma}{T M} \sum_i \sigma^x_i \right)\right]^M + O\left(\left[\frac{\Gamma}{T M}\right]^2\right)
\end{equation}
Where $H_P$ is in this paper a diagonal matrix, $\sigma^x$ are Pauli operators, and $M$ is the number of {\em Trotter slices} into which the imaginary time dimension is discretized.  We can associate the trace of the transformed quantity on the right to a sum over classical states of dimension $N \times M$ (space by imaginary time), called world-lines. For the TFIM there is no {\em sign problem}---i.e., the terms have positive weights---thereby allowing a Markov chain Monte Carlo (MCMC) integration~\cite{Evertz2003,Rieger1999,Sandvik2003}. The MCMC updates can be interpreted as a non-physical dynamics that is nevertheless known to capture correctly scaling in classical dynamics, and some special types of scaling in physical quantum dynamics~\cite{Isakov2016}.

\begin{figure}
  \includegraphics[scale=.8]{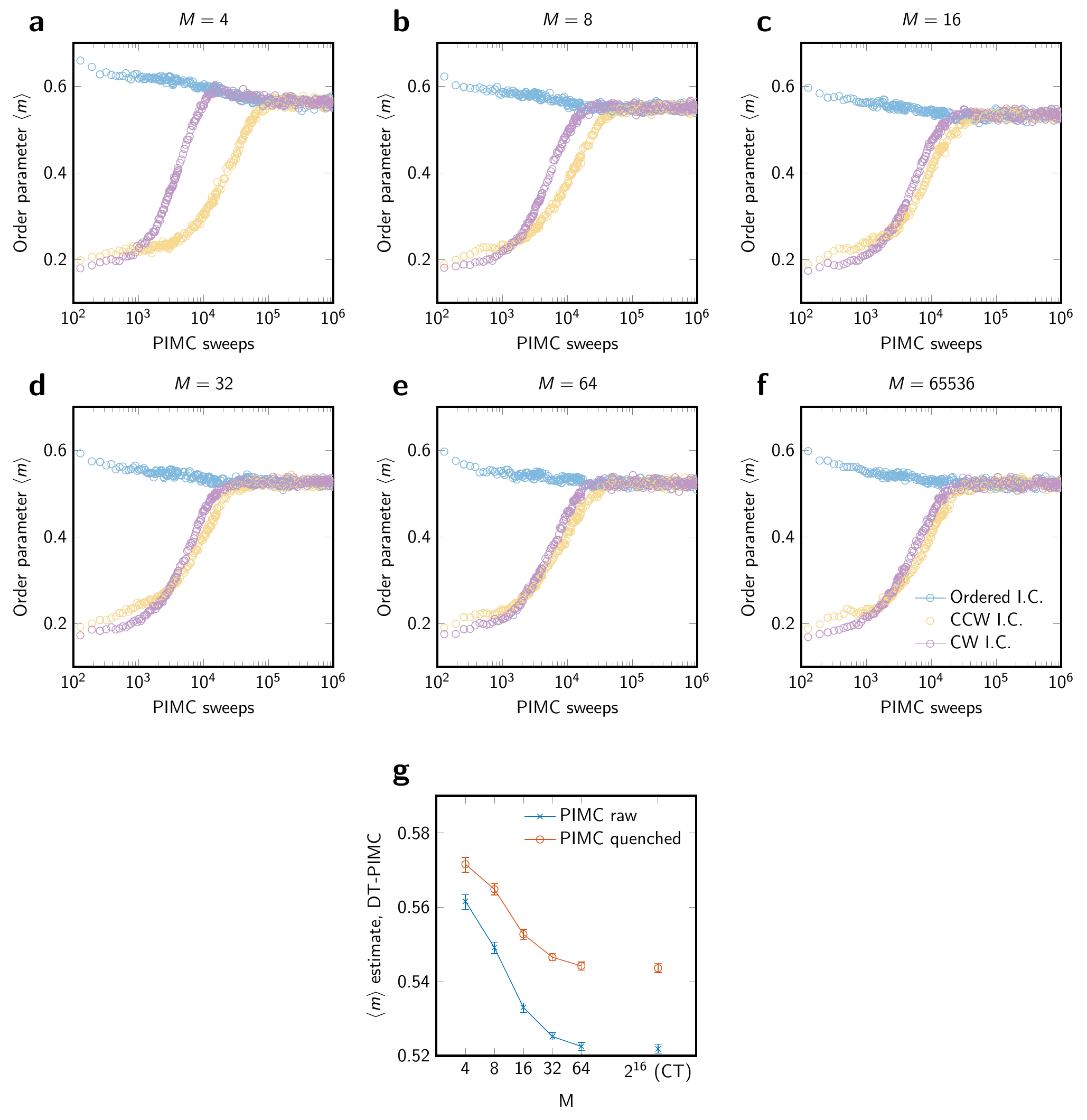}
  \caption{{\bf Trotter error in discrete-time PIMC.}  {\bf a--f},~convergence of $\braket m$ estimates, analogous to Fig.~\ref{fig:2}b, for discrete-time PIMC with varying number of Trotter slices $M$, up to the continuous-time limit $M=2^{16}$ ({\bf f}).  Model parameters are $\Gamma/J = 0.736$ and $T/J = 0.244$, with $L=15$ ($1440$ spins).  Imaginary time discretization has a strong effect on CCW/CW asymmetry in the lattice.  {\bf g}, varying $M$ leads to varying estimates of order parameter $\braket m$ at equilibrium, for both raw and quenched PIMC samples.  Estimates are taken from $2^{20}$ four-qubit update Monte Carlo sweeps; data points represent an average of $600$ replicas from each initial state; error bars are $95\%$ confidence intervals from bootstrap.
  }\label{fig:trottererror}
\end{figure}

The associated Trotter error of discrete-time PIMC (DT-PIMC) can be driven to zero by making $M T/\Gamma$ much larger than $1$; in our CT-PIMC simulations we have taken $M$ to be $2^{16}$, large enough to reflect the infinite limit over our parameter ranges.  However, one can compromise between Trotter error and convenience of algorithmic implentation; values of $M=32$ or $64$ are particularly convenient for bit-packing of the state into standard containers for which fast machine instructions exist, and allow for implementations on bandwidth limited platforms like GPUs. The methods we study all exploit the Swendsen-Wang algorithm over small qubit sets, though it is worth noting that for very small $M T/\Gamma$ other non-cluster methods may become feasible in principle.   In Fig.~\ref{fig:trottererror} we show DT-PIMC estimates of $\braket m$ as a function of the number of Trotter slices $M$.  Both convergence and equilibrium values are affected by the choice of $M$ for parameters within our experimental range.

\subsubsection{PIMC timing}

\begin{figure}
  \begin{tikzpicture}
    \node[anchor = north west] at (-.3,.2) {\includegraphics[scale = 0.8]{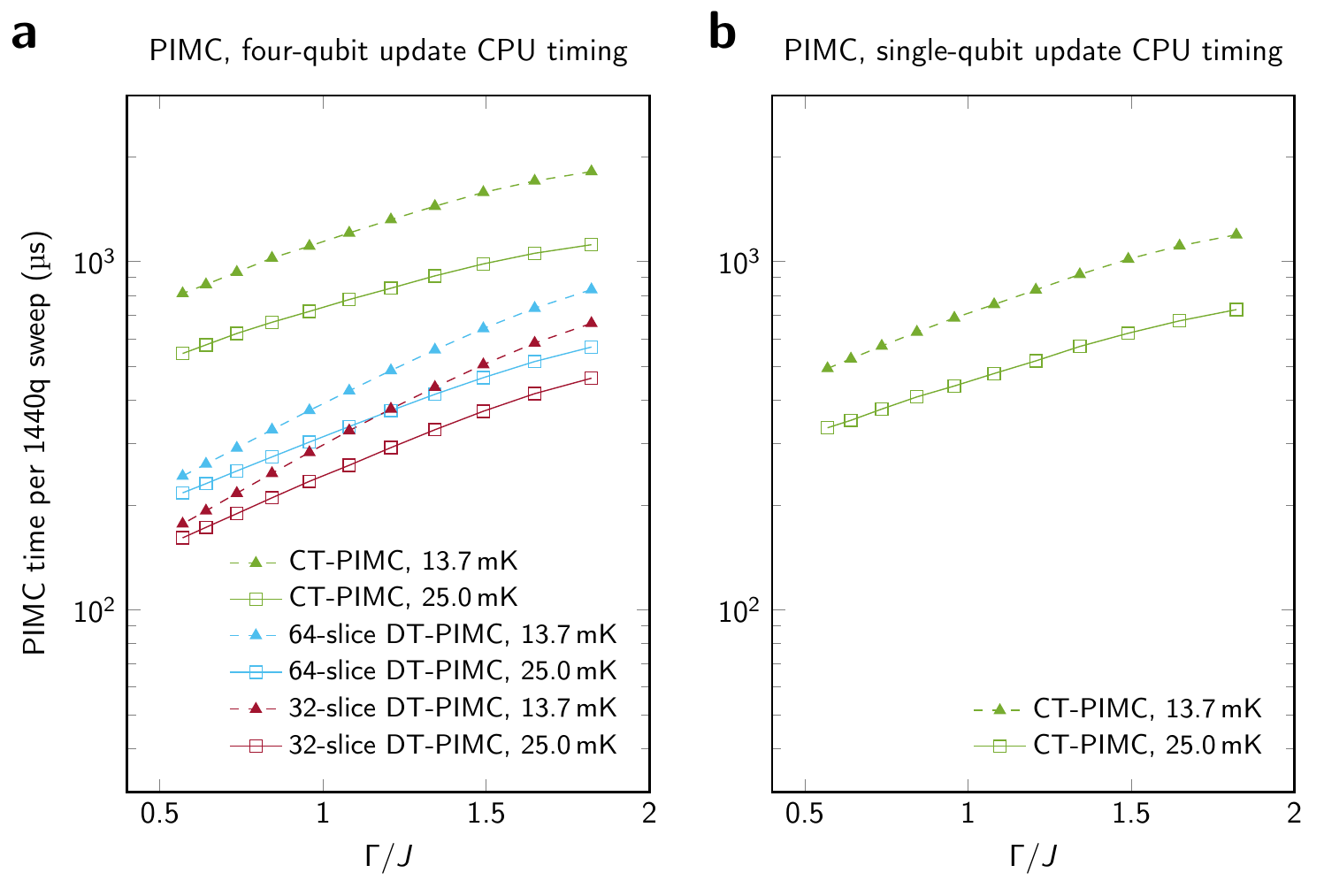}};
    \end{tikzpicture}
    \caption{{\bf PIMC timing.} Shown are average single-thread CPU time for a 1440-spin Monte Carlo sweep for {\bf a}, four-qubit updates for continuous-time and discrete-time PIMC and {\bf b}, single-qubit updates for continuous-time PIMC.  Although single-qubit updates are faster in absolute terms, convergence in the square-octagonal requires many more single-qubit sweeps than four-qubit sweeps (TODO show this).  Reported times in the paper use four-qubit continuous-time updates. 
Although time per sweep for a simple single-spin-update method  ({\bf b}) is faster than for four-qubit updates, convergence requires many more sweeps and hence the single-spin-update method is not competitive.
}\label{fig:cputiming}
\end{figure}

Fig.~\ref{fig:cputiming} shows timings of several variants of PIMC on a single CPU thread.  For each variant, sweep time is dependent on $T/J$, $\Gamma/J$, and system size.  Since time per MC sweep grows approximately linearly in the number of spins, we report only times for the largest system ($L=15$, $1440$ spins).  Time per sweep varies as a function of $T/J$ and $\Gamma/J$; we show the warmest and coldest QA temperatures over a range of $\Gamma/J$ corresponding to annealing parameter $s$ between $0.30$ and $0.40$.  CT-PIMC and DT-PIMC code was run single-threaded on two comparable CPUs: CT-PIMC on an Intel(R) Xeon(R) CPU E5-2690 v3 and DT-PIMC on an Intel(R) Xeon(R) Platinum 8275CL, respectively.

We also measured timings for 32-slice, four-qubit-update DT-PIMC on an NVIDIA Tesla V100 GPU.  For the 1440-spin lattice, running a single experiment gives a per-sweep time of roughly $\SI{40}{\micro s}$ independent of $T$ and $\Gamma/J$.  The experiment can be run on a GPU with a tradeoff between latency and throughput; optimizing for total throughput increases latency to roughly $\SI{50}{\micro s}$ per sweep, but decreases the total time needed for an ideally structured experiment by a factor of roughy $120$.  This means that for an appropriately structured experiment, the high-end GPU can provide results approximately $1000$ times faster than a single CPU thread for 32-slice DT-PIMC.

\begin{figure}\includegraphics[scale = 0.75]{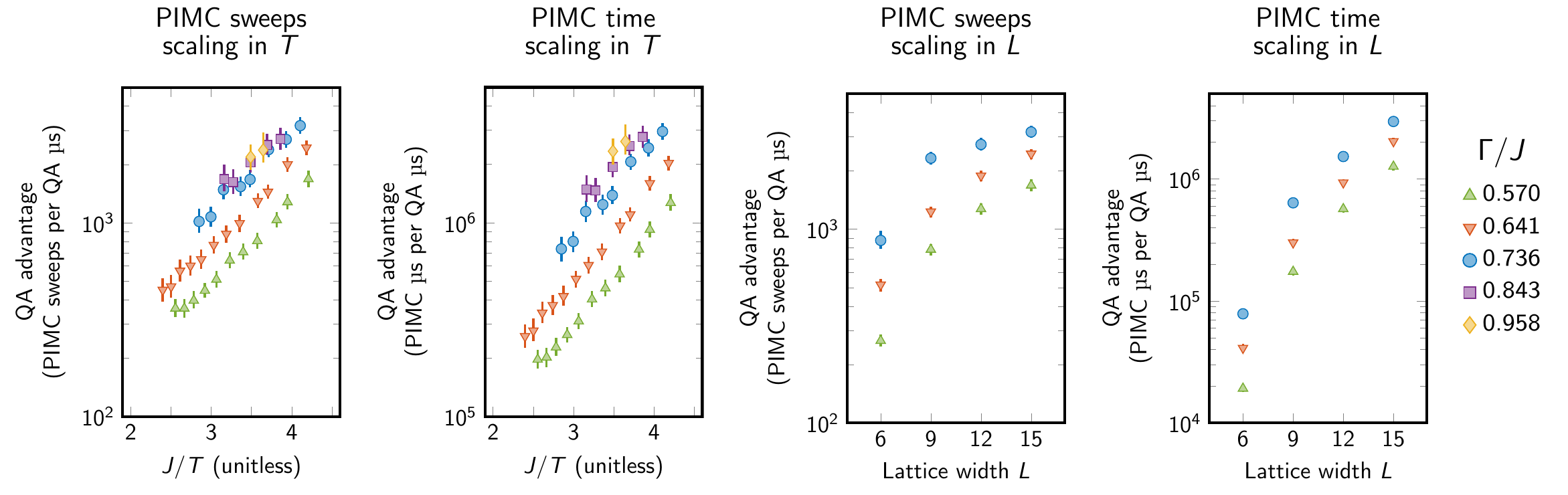}
  \caption{{\bf Scaling of QA against PIMC.}  Ratio of QA and PIMC time, measured either in PIMC sweeps or in CPU microseconds.
}\label{fig:qmc_extras}
\end{figure}

Fig.~\ref{fig:qmc_extras} shows PIMC convergence time as a function of QA convergence time, as in Fig.~\ref{fig:4}, in units of both PIMC sweeps and PIMC CPU time.

\subsubsection{Wall-clock time and power usage}

We have compared convergence timescales in QA and PIMC simulations with the aim of studying relaxation dynamics.  To take more practical benchmarking perspective, one can consider the total time taken to generate QA output samples, which we have made little effort to optimize.  The QEMC cycle is dominated by a $\SI{10}{ms}$ wait time between consecutive samples; this parameter was chosen conservatively to allow settling of the system and to minimize heating.  A spot-check indicated that reducing this to $\SI{2}{ms}$ resulted in no statistically significant difference in behavior. The next dominant factors are a programming time of $\SI{10}{ms}$ (amortized over the number of samples drawn) and a per-sample readout time of $\SI{274}{\micro s}$.  As a result of the wait time, the wall clock time of the QA processor is roughly $10^4$ times longer than the relaxation time under study.  Removing the wait time and drawing many samples per programming leaves the QA duty cycle dominated by readout time, which in the case of $t_p=\SI{4}{\micro s}$ renders the wall clock time roughly $10^2$ times longer than the relaxation time.  Since the focus of the study was relaxation dynamics rather than optimization of wall-clock time, we have not probed the effect of further reducing the wait time.

One important consideration in the utility of quantum processors is energy consumption.  The QPU apparatus draws $\SI{25}{KW}$, dominated by refrigeration.  The 8176 and E5-2690 CPUs draw $\SI{6}{W}$ and $\SI{11}{W}$ per core respectively;

The i7-8650U CPU has a power specification (TDP) of $\SI{15}{W}$; we can estimate the power-per-core (4 cores) as $\SI{3.75}{W}$.  The V100 GPU has a power specification of $\SI{300}{W}$ and, as mentioned above, can be up to $1000$ times faster than a CPU core for this particular application.  Thus taking the maximum time-advantage of QA over a CPU thread and an entire V100 GPU as approximately $10^6$ and $10^3$ respectively, we see that the quantum processor---counting only the relaxation pause time---can be over an order of magnitude more power-efficient than both CPU and GPU.  If we consider QA wall-clock time, as discussed above, we see no power advantage.  However, we expect the scaling of power requirements for near-term QA processors to show very flat scaling compared to computational power, since the dominant power draw---from a pulse-tube dilution refrigerator---is fairly constant.

\subsection{Effect of quench and disorder}

\begin{figure}
  \begin{tikzpicture}
    \node[anchor = north west] at (-.3,.2) {\includegraphics[scale = 0.8]{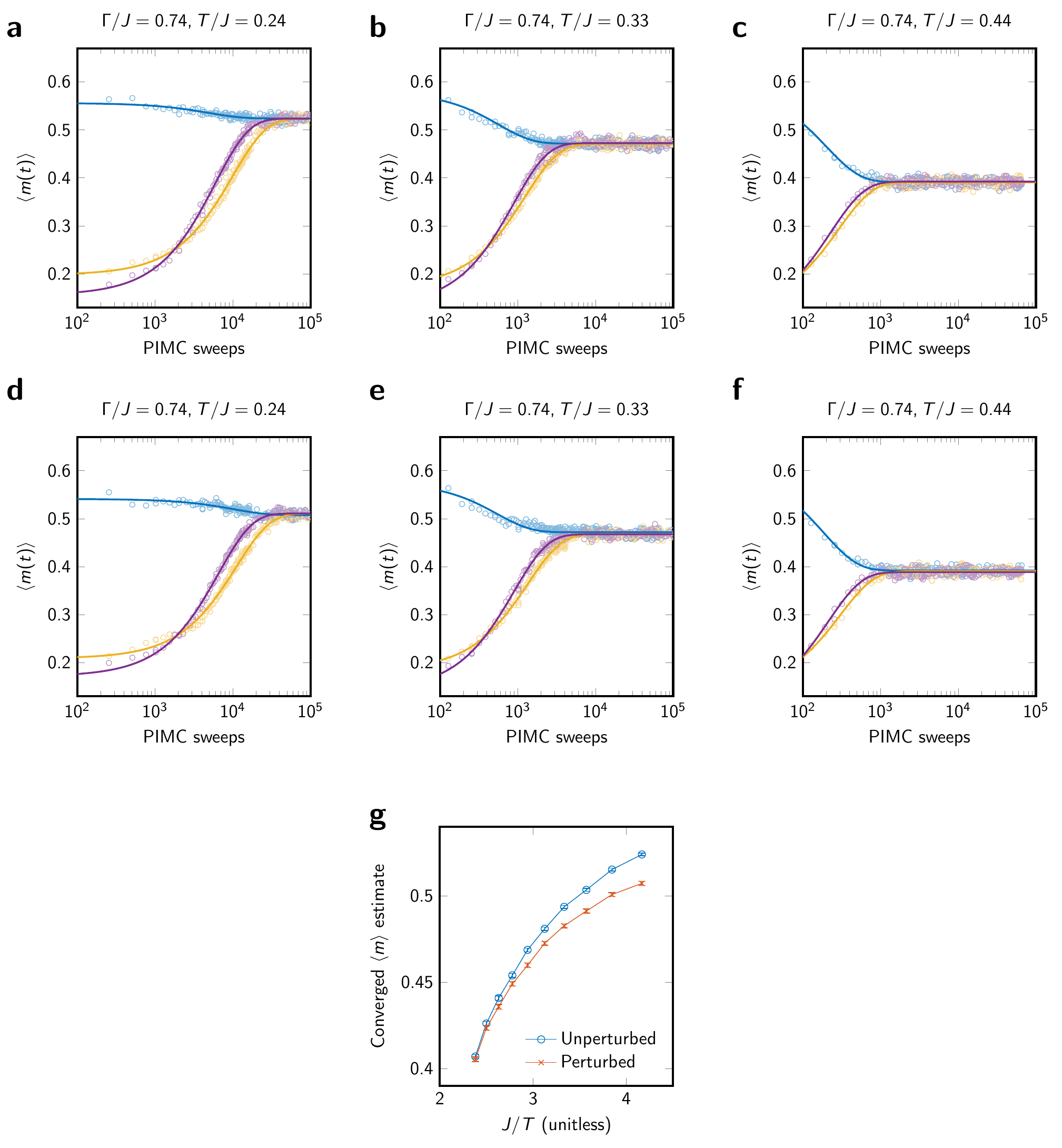}};
    \end{tikzpicture}
    \caption{{\bf Effect of perturbations in the Ising Hamiltonian.} We show projected PIMC measurements as they converge in the unperturbed Hamiltonian ({\bf a--c}) and with quenched disorder (perturbation of terms with $\sigma=0.02$) added to the Ising Hamiltonian ({\bf d--f}).  Equilibrium estimates of $\braket m$ show that disorder suppresses $\braket m$, with a temperature-dependent effect that is largest at low temperatures.  This is consistent with QA data and demonstrates a disordering effect that does not accelerate convergence.
}\label{fig:ice}
\end{figure}

Two important differences between the QA processor and an ideal system are effects of the readout quench, and analog misspecification in the Hamiltonian, e.g., device inhomogeneity.  To understand the effect of this misspecification on statistical estimates, we run PIMC with static (quenched) disorder in the Hamiltonian.  We perturb the classical Ising Hamiltonian by adding i.i.d.\ Gaussian terms with standard deviation $\sigma=0.02$ to each linear term $h_i$ and to each nonzero coupling term $J_{ij}$.  We instantiate these errors independently for each sample, giving 600 instantiations of error for each initial condition.  Results are shown in Fig.~\ref{fig:ice}.  The perturbations in the Hamiltonian have no significant effect at high temperatures, but suppress $\braket m$ slightly at low temperatures; this is consistent with Fig.~\ref{fig:2}c, where we see QA estimates deviating slightly below quenched PIMC estimates at low temperatures.  The analog error does not change convergence timescales significantly.

\begin{figure}
  \begin{tikzpicture}
    \node[anchor = north west] at (-.3,.2) {\includegraphics[scale = 0.83]{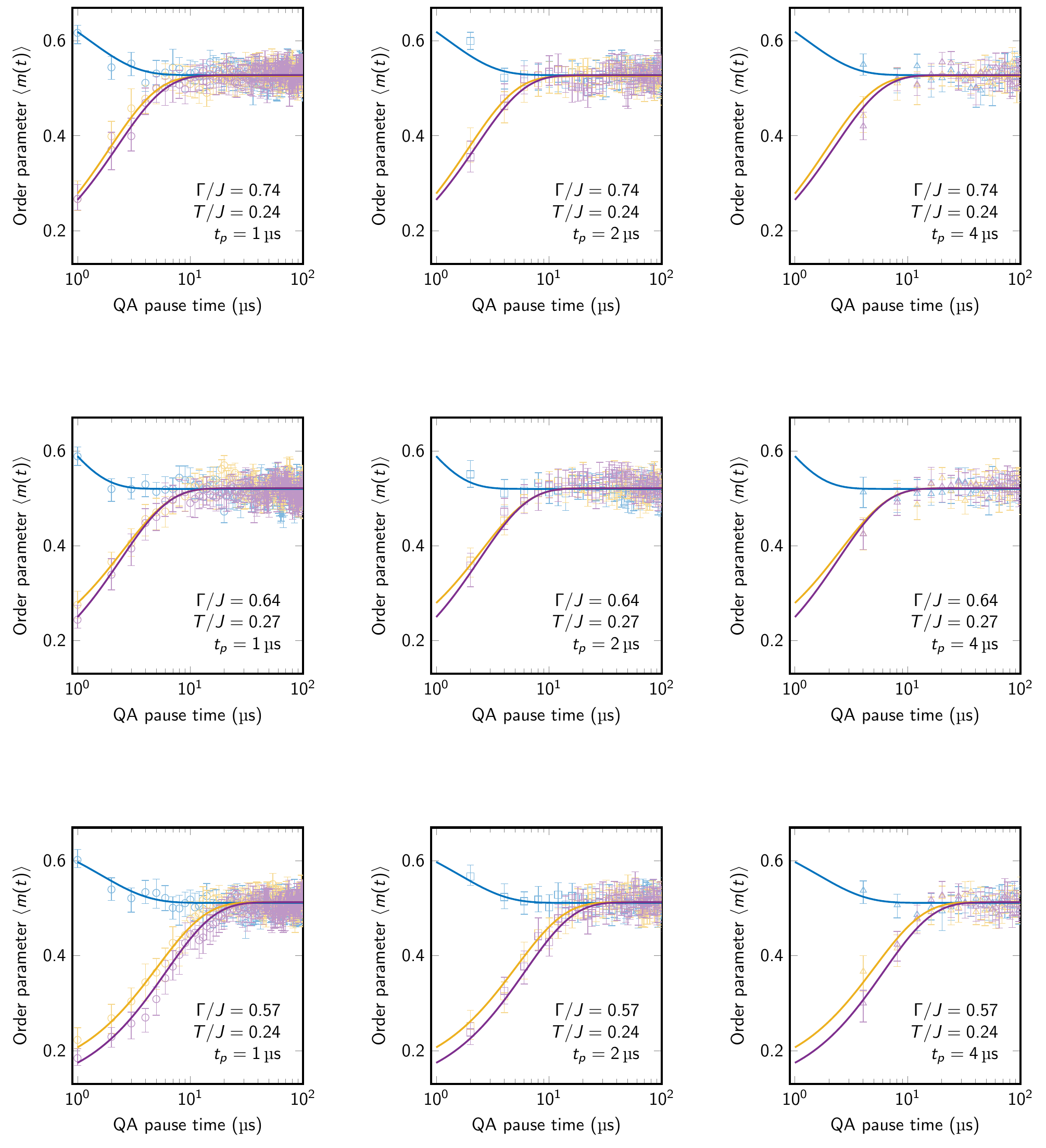}};
    \end{tikzpicture}
    \caption{{\bf QA convergence for varying parameters and pause times.} We show QA measurements as in Fig.~\ref{fig:2}b with different values of pause time $t_p$ separated.  Each row is a different value of $(\Gamma/J,T/J)$.  For each row, the same exponential fit is shown.
}\label{fig:qa_convergence_tp}
\end{figure}

Fig.~\ref{fig:quenchsensitivity} demonstrates that changing the length of the quench has little systematic impact on relaxation timescales.  That is, although the total length of the quench and reverse anneal are similar to the pause time $t_p$, dynamics appear to be frozen almost immediately.  However, this does not rule out the possibility that the act of repeatedly quenching and reverse annealing in QA distorts the measured timescales.  Fig.~\ref{fig:qa_convergence_tp} shows QA data for several slow-converging models at $t_p \in \{ \SI{1}{\micro s},  \SI{2}{\micro s},  \SI{4}{\micro s} \}$.  We see no evidence of systematic distortion of the data arising from the quench protocol.

We now consider effects of the readout quench on estimated equilibrium statistics.  In this experiment the QA readout quench has been used as an approximation to projective readout.  However, the system shows evidence of local relaxation during this quench; we estimate that the timescale of relevant dynamics is on the order of tens of nanoseconds.  To address this issue, we take two approaches.

The first approach is to quench PIMC output, which we do by taking a PIMC spin state, projected to the $\sigma^z$ basis, and applying a greedy descent in the classical potential: first we repair frustrated four-qubit chains by majority vote, breaking ties randomly.  Next we greedily flip chains while doing so lowers the classical energy.  This allows us to compare quenched PIMC output with quenched QA output; this is the approach we take in the main body of this work.

The second approach is to ``unquench'' QA output by applying a small number (10) of PIMC postprocessing sweeps.  After this postprocessing we project the PIMC state to the $\sigma^z$ basis as with our PIMC experiments; this allows us to compare unquenched QA output with projected PIMC output.

\begin{figure}
  \includegraphics[scale=.8]{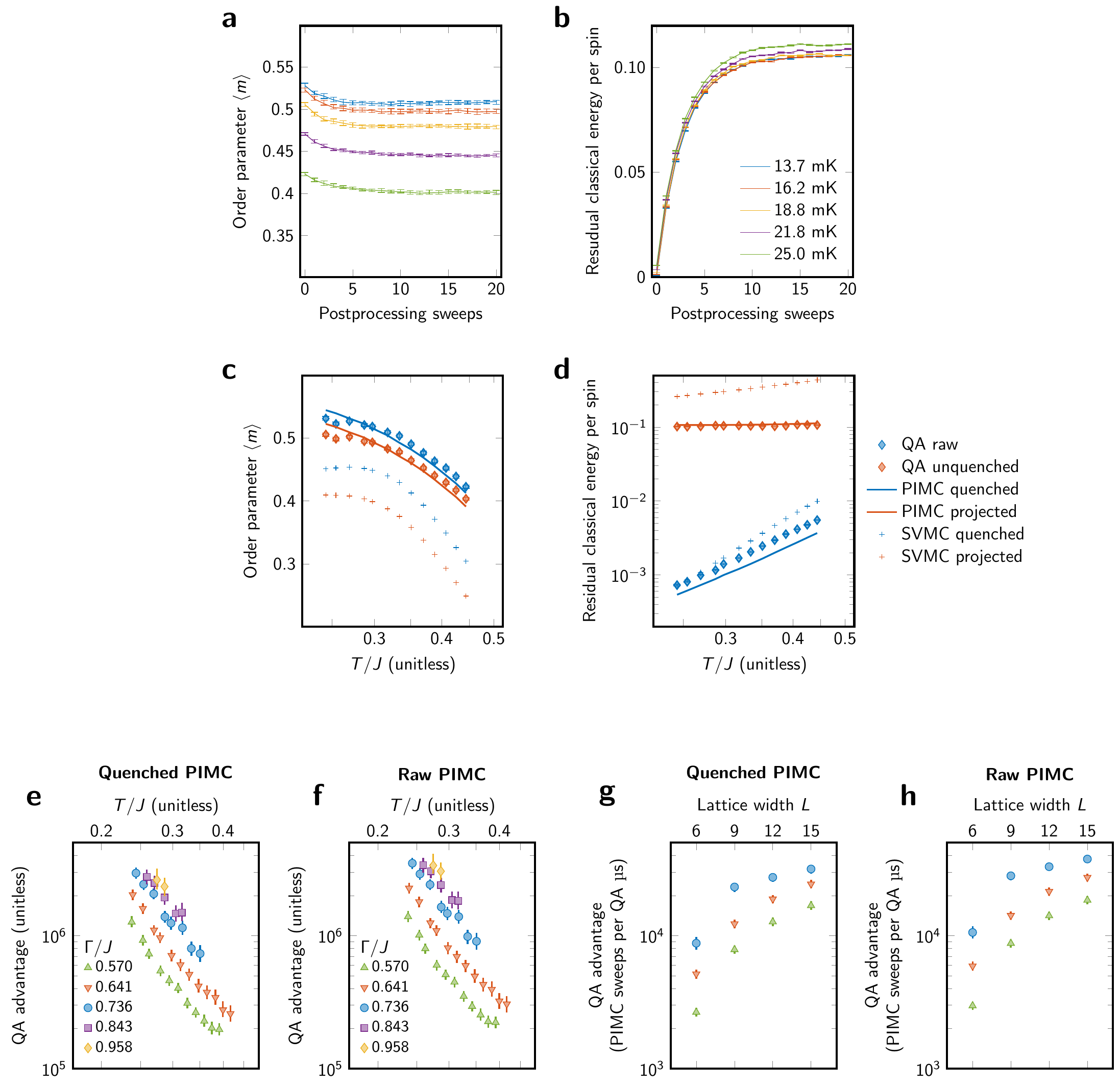}
  \caption{{\bf Quenching and unquenching samples.} {\bf a--b}, To mimic a reversal of the QA readout quench, we apply a small number of PIMC postprocessing sweeps to QA output samples.
    {\bf a}, The effect on $\braket m$ is modest, and similar in magnitude to the effect of a classical quench on PIMC output samples.
    {\bf b}, The effect on classical residual energy is significant and is mostly complete after $10$ sweeps.
    {\bf c--d}, Quenching PIMC output and unquenching QA output (with 10 sweeps) allows us to compare quenched and unquenched statistics for a range of temperatures with $\Gamma/J = 0.736$.  We compare QA results to both PIMC and SVMC (a semiclassical rotor model). {\bf c}, In both QA and PIMC, quenched estimates of $\braket m$ are roughly $0.02$ higher than unquenched estimates.  QA and PIMC agree across the range of parameters; SVMC deviates significantly.  {\bf d}, Residual classical energy differs dramatically between quenched and unquenched samples in all models, indicating its sensitivity to local processes and therefore the short timescale of the physical QA quench.
  {\bf e--h},~ We use quenched and raw PIMC samples to reproduce Fig.~\ref{fig:4}c ({\bf e--f}) and Fig.~\ref{fig:4}d ({\bf g--h}); our main conclusions regarding QA and PIMC relaxation rates are not affected by whether or not we quench PIMC samples.}\label{fig:sm_postprocess}
\end{figure}

In Fig.~\ref{fig:sm_postprocess} we show the effect of this unquenching on the average order parameter and residual energy.  While the order parameter is relatively robust to the process due to being somewhat topologically protected, the PIMC postprocessing sweeps dramatically increase the residual classical energy, which is dominated by local excitations (i.e., bound vortex-antivortex pairs).  We see that raw QA output has systematically higher residual energy than quenched PIMC output; this is because excitations arising from frustrated ferromagnetic bonds are completely removed by the PIMC classical quench.  We confirm in Fig.~\ref{fig:sm_postprocess}e--h that our conclusions about advantage of QA over PIMC does not depend on the quench.

\subsubsection{Effect of quench on relaxation timescales}

\begin{figure}
  \includegraphics[scale=.8]{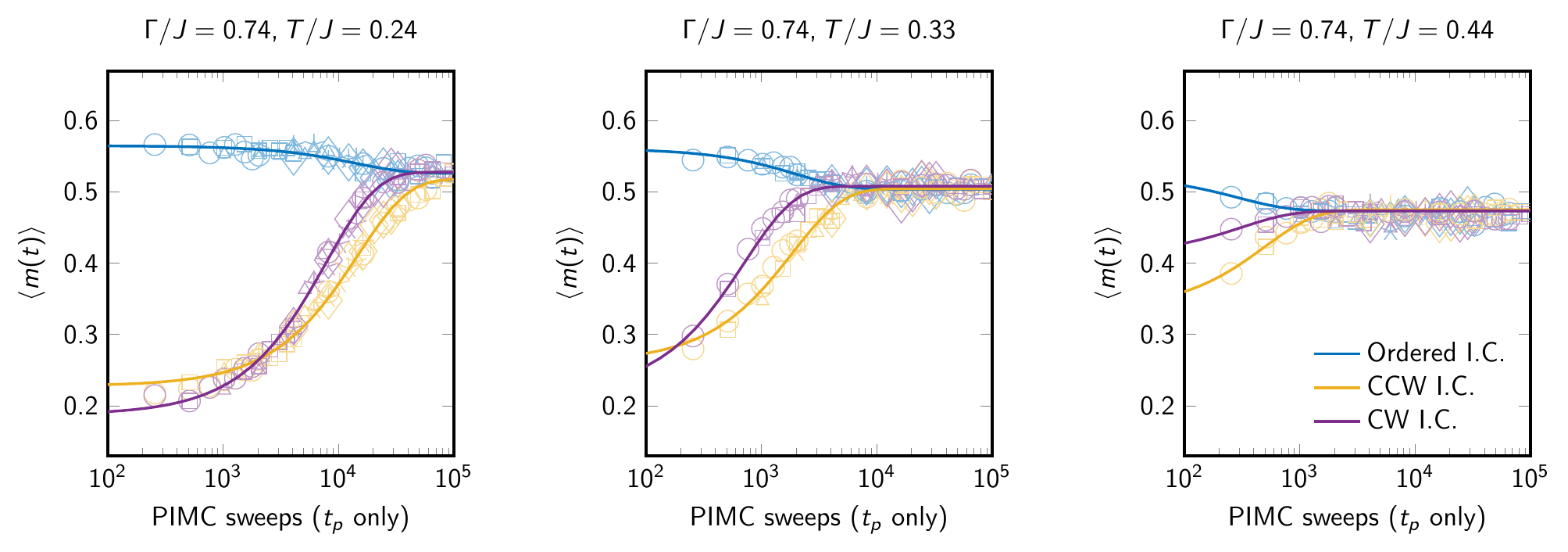}
  \caption{{\bf Modeling QA quench in PIMC.} We assign $t_q$ to be 1000 PIMC sweeps, estimating an equivalent rate from QA data.  We then vary $t_p$ in PIMC sweeps (see Fig.~\ref{fig:quenchsensitivity}).  Different markers indicate different pause lengths $t_p$; we perform eight steps of each length.}
    \label{fig:fullquench}
\end{figure}

The PIMC quench we have used up to this point is a simple one-way postprocessing that does not influence the dynamics of PIMC itself.  Figs.~\ref{fig:quenchsensitivity} and \ref{fig:qa_convergence_tp} show that QA convergence timescales are not strongly affected by the QA quench protocol, so it is appropriate to compare QA dynamics against the dynamics of fixed-Hamiltonian PIMC, only using the quench to roughly quantify the effect of local relaxation.  However, the effect of quenching in PIMC is still interesting.  To investigate another model of PIMC quench that can potentially affect dynamics, we assume an equivalent PIMC-to-QA rate of 1000 sweeps per QA microsecond.  This is a reasonable approximate value (Fig.~\ref{fig:4}c), and allows us to model the full QA protocol of reverse annealing and quenching.  Results are shown in Fig.~\ref{fig:fullquench}.  Adopting this more complex PIMC quench does not provide a speedup; the difference in convergence time between $T/J=0.24$ and $T/J = 0.33$ remains just under an order of magnitude (cf.~Fig.~\ref{fig:4}a).  In addition, $\braket m$ is overestimated far more at $T/J = 0.44$ than in QA.

\begin{figure}
  \includegraphics[scale=.8]{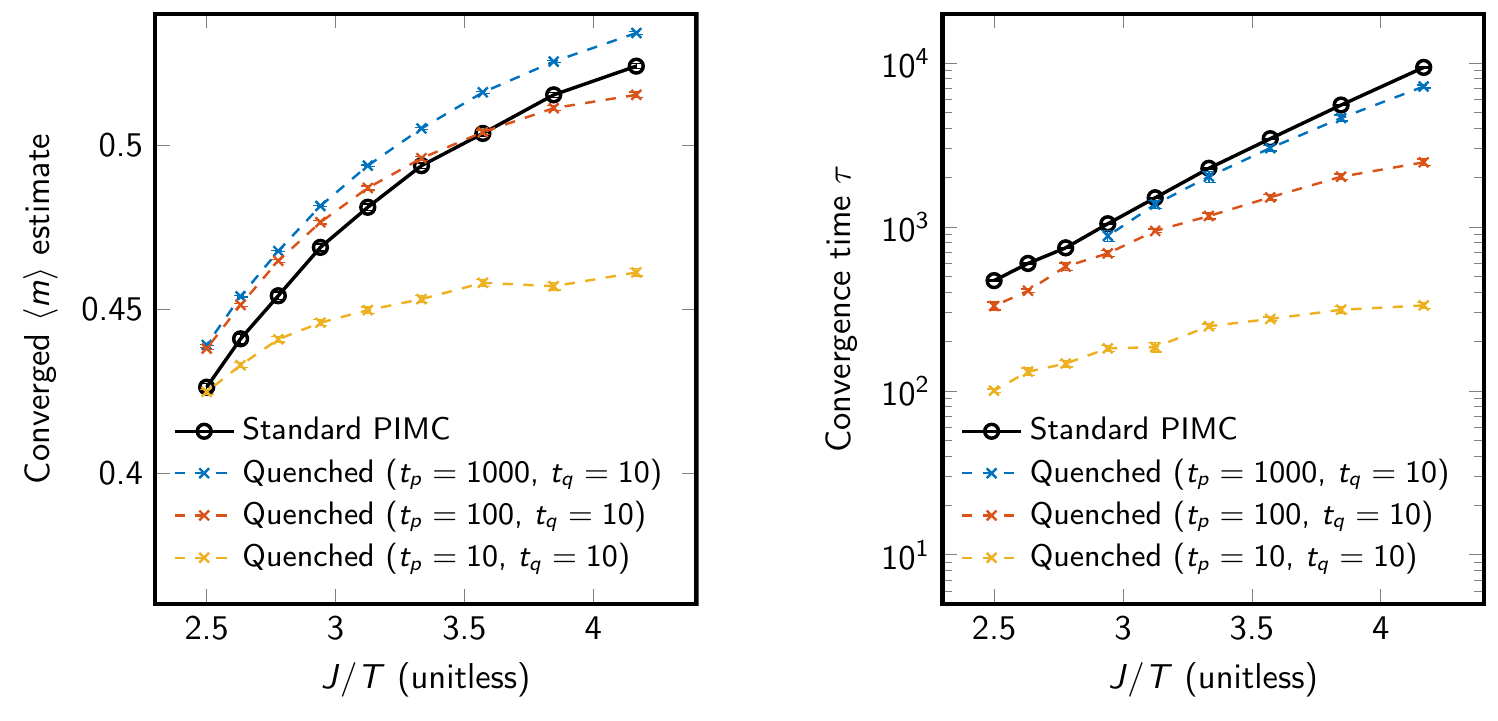}
  \caption{{\bf Effect of alternative quench methods in PIMC.} Fig.~\ref{fig:fullquench} modeled QA quench in PIMC using approximately corresponding, physically motivated parameters.  Here we explore an alternative approach, quenching only $\Gamma/J$ instead of both $T/J$ and $\Gamma/J$, and we do so more frequently and quickly than corresponding QA parameters.  At $\Gamma=0$ and $T>0$ the system, like the triangular Ising antiferromagnet, has no long range order.  Consequently at low temperatures there is a temperature-dependent effect on order (left) and acceleration of escape (right).  Quench frequency provides a tradeoff between speed of escape and simulation accuracy.  Error bars are from exponential fitting.}
    \label{fig:pimcfastquench}
\end{figure}

We also explore an alternative PIMC quench method that differs from the QA protocol.  Instead of quenching both thermal and quantum fluctuations as in QA (and Fig.~\ref{fig:fullquench}), we quench only $\Gamma$, linearly over a few PIMC sweeps.  We perform this quench with free parameters $t_p$ and $t_q$ rather than with parameters motivated by the QA protocol.  Fig.~\ref{fig:pimcfastquench} shows data in which each sample is derived from the previous by $\Gamma$-only reverse anneal and quench of duration $t_q=10$ sweeps, with a pause of $t_p\in \{10,100,1000\}$ sweeps.  Note that this is more frequent than QA quench in our experiments.  The data shows that fast quenches in PIMC can lead to a highly non-equilibrium dynamics and can provide speedup.  Data is qualitatively similar for $t_q=0$ sweeps, where PIMC is periodically projected to the classical space.  Pause length $t_p$ provides a tradeoff between escape dynamics and simulation accuracy; for $t_p=100$ the effect on escape dynamics is large compared to the effect on $\braket m$.  We hypothesize that the dynamical effect is related to the fact that when $\Gamma=0$ and $T>0$, the effective Hamiltonian for four-qubit PIMC updates is a classical triangular Ising antiferromagnet, which has no order-by-disorder effect compelling neighboring pseudospins to align exactly.

\subsection{Local signatures of entanglement}\label{sec:entanglement}

Entanglement is a property restricted to quantum systems. At finite temperatures the equilibrium distribution of a quantum system is described by a mixed state, and this makes unambiguous establishment of entanglement difficult. Here we examine two entanglement witnesses: bipartite concurrence~\cite{Hill1997} and the Peres-Horodecki criterion~\cite{Peres1996,Horodecki1996} (positive partial transpose), both of which demonstrate interesting forms of entanglement that place limitations on the ability of quasi-classical models to simulate the system. Significant entanglement is present at the temperature and transverse field ranges examined; these simple measures indicate increased entanglement with decreasing temperature over the experimental range.  This provides a plausible explanation---increasing entanglement---for the increasing computational advantage seen in the quantum hardware as temperature decreases.

\begin{figure}
  \includegraphics[scale=0.8]{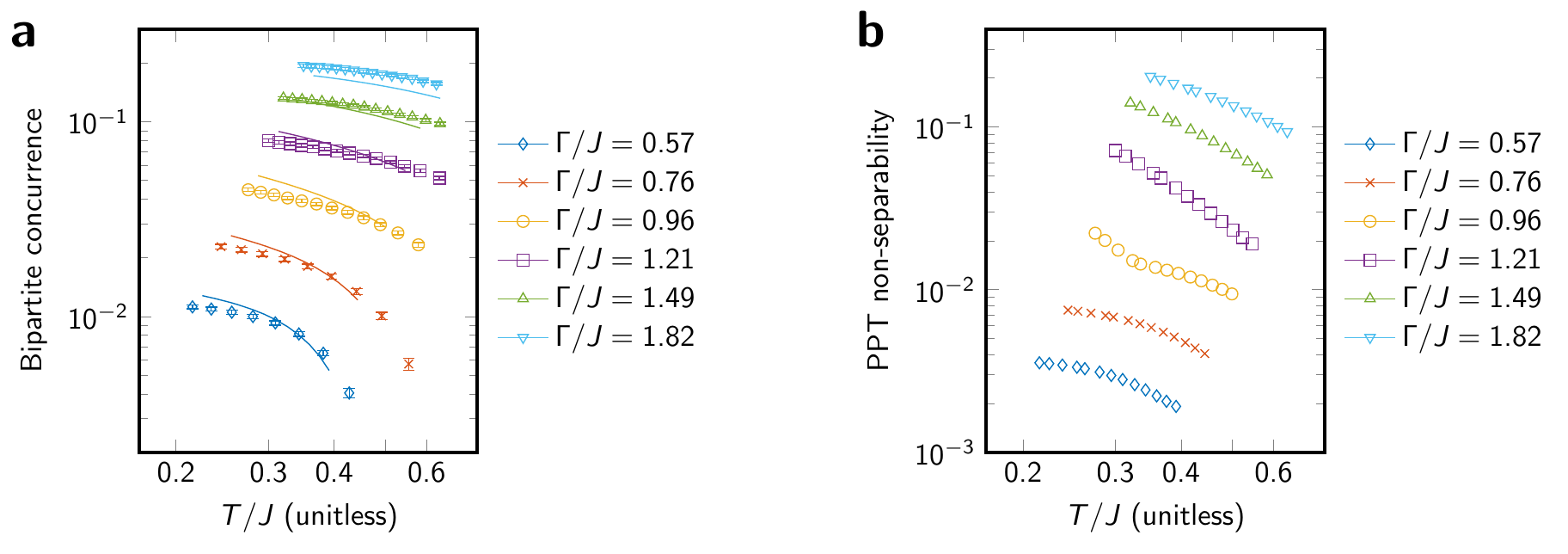}
  \caption{{\bf Entanglement witnesses.}
    {\bf a},~We consider bipartite concurrence as a witness of entanglement between two central spins in a four-qubit chain.  Lines indicate exact diagionalization results on the 12-qubit gadget (Fig.~\ref{fig:squids}a), and marks indicate estimates from quantum Monte Carlo on central chains in a cylndrical lattice with $L=6$.  Concurrence increases with $1/T$ and with $\Gamma$ in the experimental parameter range.
  {\bf b},~By considering the smallest eigenvalue (most negative) associated to the 12-qubit gadget instance under partial transpose of one 4-qubit block, we can establish a witness for entanglement between chains. The absolute value of the negative eigenvalue is plotted; as expected it decreases with increasing temperature, and peaks at intermediate transverse field, indicating the failure of factorized approximations (mapping four-qubit chains to logical spins) throughout the experimental range. This measure is limited to the 12-qubit gadget due to the computational complexity of the witness on larger systems.}\label{fig:entanglement} 
\end{figure}

Certain forms of bipartite entanglement can also be measured in large systems by a quantum Monte Carlo method.  Many terms in the density matrix can be estimated using standard PIMC; measurement of certain off-diagonal terms---as a ratio of partition functions---requires modification of the imaginary time boundary conditions.  By doing so we can measure traced density matrices in large systems and establish bipartite concurrence in our larger lattices.  Bipartite entanglement results for neighboring central spins on four-qubit chains are shown in Figure \ref{fig:entanglement}a. We find strong concurrence at the level of pairs across the experimental parameter range. We show this both in the 12 qubit gadget system, and in the $L=6$ lattice under investigation. This result contrasts with the perturbative argument for the nature of entanglement, which is based only on the presence of GHZ states.  The presence of entanglement between two qubits in a chain, after tracing out the other two, indicates rich quantum correlations beyond the perturbative argument.  Thus we establish not only concurrence, but that its form is notably distinct from that predicted in the perturbative regime across the parameters studied.

We also applied this entanglement measure to qubits beyond chains at various distances, but did not find an unambiguous signal for quantum correlations. Tracing out of variables is necessary for a practical method at scale based on this entanglement witness, but in the process evidence of entanglement can be destroyed. However, we can use alternative witnesses over small systems. The absence of entanglement allows for factorization of the distribution with respect to chains. To test this we can use the Peres-Horodecki criterion at the level of a chain; a partial transpose with respect to the chain will yield no negative eigenvalues if there exists a separable (quasi-classical) approximation. As shown in Figure \ref{fig:entanglement}b, this criterion indicates the presence of entanglement beyond the chain level throughout the experimental range, with the evidence becoming weaker at larger temperature and smaller $\Gamma$ as expected.

It should be noted that the bipartite concurrence and Peres-Horodecki witnesses chosen to demonstrate entanglement here are sufficient, but not necessary, for the demonstration of entanglement and we have not been able to link them to a specific accelerating dynamical mechanism. Developing and applying witnesses better tailored to the lattice structure and phases, and exploiting only QA-accessible information, is future work.

\subsection{Simulation via approximate models}

Fully-frustrated two-dimensional transverse field Ising models are known to exhibit a number of interesting behaviors, not least the order-by-disorder phenomenon and competition between quantum and thermal fluctuations~\cite{Jalabert1991,Moessner2000,Moessner2001,Isakov2003}.  Interesting scaling has been exhibited in finite-size simulations of the square-octagonal and triangular lattices, attributed in part to a crossover with quantum critical phenomena~\cite{Isakov2003,King2018}. A simple exact description of the pertubative limit (small $T$ and small $\Gamma$) for the square-octagonal lattices demonstrates the importance of four-qubit GHZ states in the ground state~\cite{King2018}.  In this paper we have examined the dynamics of twisting; this is intuitively related to critical behavior in the form of system-spanning vortices.  We have demonstrated results that indicate dynamical differentiation between world-line dynamics (of PIMC) and physical QA dynamics. These various factors indicate the importance of quantum phenomena in describing the equilibrium properties of the lattice in question.

PIMC is a standard tool for simulating equilibrium properties of a finite-temperature TFIM.  However, we want to confirm that we have compared QA against the best classical simulation in the specific setting of the square-octagonal lattice.  Thus we need to rule out alternative approaches that could potentially simulate the square-octagonal lattice more efficiently than PIMC.  The entanglement related to the GHZ states in the perturbative limit is localized.  Near this limit, properties of the phase diagram can be explained without long-range quantum correlations.  Furthermore the KT phase, being a finite-temperature critical phase, is known to have classical scaling: the scaling of extensive properties in system size near the critical point is consistent with classical models at large scale\cite{Herbut2007}.
Finally, the low energy solution space seems to indicate the possibility for a reduction of the model from a square-octagonal to a triangular description by mapping four-qubit chains onto logical qubits, or an even simpler phenomenological model over plaquette pseudospins.
Given these factors it is reasonable to ask if there are approximations to the square-octagonal transverse field Ising model that would allow for equally good approximation at lower cost, or expose the absence of quantum dynamics in accelerating unwinding.

Although it is impossible to disprove the existence of a superior simulation method for the lattice in question, in this section we rule out several reasonable approaches: quasi-classical approximations consistent with the absence of entanglement and quantum mechanisms, 
reductions of the square-octagonal model to a triangular lattice model,
and a phenomenologically matched XY model.  We have already demonstrated evidence for entanglement at short range, which places some restrictions on the quality of approximations achievable by quasi-classical approximations.

\subsubsection{Six-state clock and other phenomenological approximations}

One potential way to bypass long equilibration timescales in the lattice under study is to replace the TFIM with a simpler classical model exhibiting the correct phases and symmetries.  The most natural candidate is the plaquette pseudospin, which accurately reflects the physics of the TFIM in the perturbative limit.  A pseudospin describes the state of a frustrated plaquette, and takes one of six values in any classical ground state $\theta_i \in \{2 \pi k/6: k=0,\ldots,5\}$. These plaquette states are constrained through shared qubits; these constraints favor full or partial alignment of neighboring pseudospins. This interaction can be qualitatively captured by pairwise ferromagnetic couplings between six-state XY spins over a dual (honeycomb) lattice
$$H = -\sum_{ij} \cos(\theta_i - \theta_j),$$
where we can determine strength of coupling by the temperature $T$. The translation from the full model interaction to this simpler pairwise one is nontrivial at the microscopic level, but $T$ can be tuned to optimize the approximation. The six-state clock model exhibits both a P-KT transition and a low temperature crystalline state\cite{Jose1977}, and retains the symmetries of the square-octagonal model with respect to the classical ground states.

The TFIM phase diagram includes a quantum critical point at $\Gamma=\Gamma_c$; the clock model cannot reproduce the QPT or the domed phase diagram of the TFIM\cite{Isakov2003,King2018}.  Still, suppose we wish to use the six-state clock model to approximate behavior in the vicinity of the P-KT phase transition for some transverse field $\Gamma$ with $0<\Gamma<\Gamma_c$.  The known critical scaling of both models dictates a divergence of the correlation length approaching the critical phase as $\xi \propto \exp(a/\sqrt{T-T_c})$.  To correctly model the susceptibility, order parameter, and other extensive quantities---in a scalable sense---it is necessary for this correlation length to be matched in both models. Knowing only the (non-universal) parameters $a$ and $T_c$ for each model we can require that the unitless correlation lengths match up to a prefactor.  With this mapping between the model parameters in place, we can examine the approximation achieved by each at a given correlation length. This amounts to a comparison of the collapse form.

\begin{figure}
\includegraphics[scale =0.8]{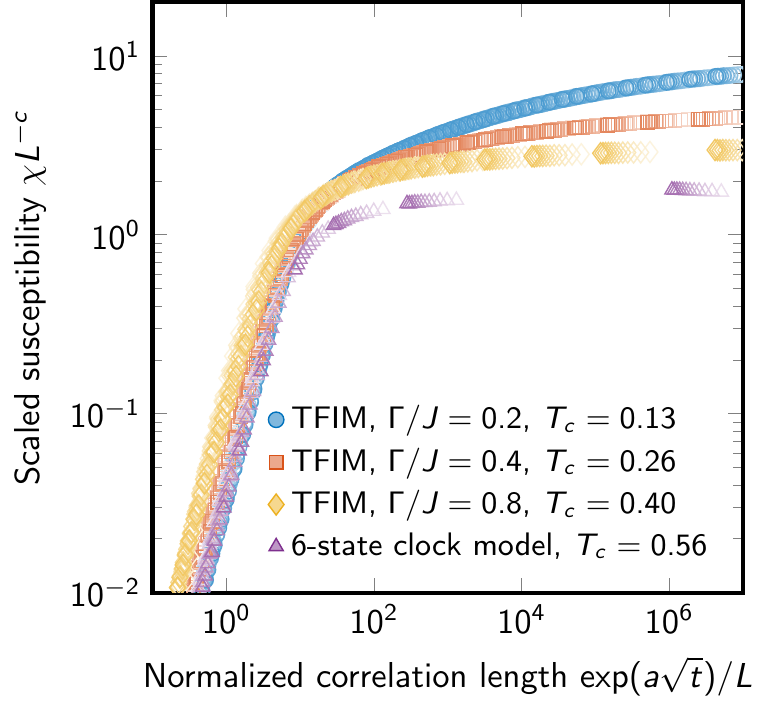}
\caption{{\bf Finite size scaling collapse of TFIM and pseudospin models.} We compare
 collapses of the triangular AFM TFIM at several values of $\Gamma/J$ with the six-state clock model derived from the plaquette pseudospin.  Scaling parameter $c$ is clamped to the value $7/4$ derived from 2D XY unversality. Marker opacity indicates system size from $L=9$ to $L=36$.  Differing scaling forms indicate that behavior of the fully-frustrated TFIM cannot be derived merely from critical temperature $T_c$, and that the classical six-state clock model fails to predict properties of the quantum system when reparameterized to match correlation length.}\label{fig:6stateclock}
\end{figure}

The six-state clock model on the hexagonal lattice can be considered an approximation to both the triangular antiferromagnet and the square-octagonal lattice.  We first show deviations between the clock model and the triangular AFM via finite size scaling collapse, then show deviations between the triangular AFM and the square-octagonal model.  To minimize finite size effects we work with periodic boundary conditions on $L \times L$ lattices.  We fitted scaling collapse parameters using a smoothness condition, fixing the parameter $c$ to $7/4$, reflecting the 2D XY universal critical exponent of $\eta = 1/4$ (Ref.~\onlinecite{Isakov2003}).  Results are shown in Fig.~\ref{fig:6stateclock}.  The difference in forms is an indication that approaching the phase transition at equivalent correlation length there are substantial differences in susceptibility of the lattice, susceptibility growing more quickly in the TFIM and to a larger peak value at the phase transition---this is beside the $\Gamma$-dependence of TFIM scaling, which the six-state clock model cannot capture.  The universal value leads to an excellent collapse of the 6-state clock model at these scales, but not so for our model due to more interesting finite size effects and a previously noted crossover impact from quantum critical phenomena\cite{Isakov2003}. Thus even after careful tuning of the parameters to match the growth of the correlation length, the predictive power of the pseudospin model for the TFIM is poor.

\subsubsection{Triangular lattice transverse field Ising model approximation}

There is a simple local one-to-one mapping between the classical ground states of the square octagonal model and those of the fully frustrated triangular model. The equivalence is realized in mapping the states of every four-qubit ferromagnetically-coupled cluster in the square octagonal lattice to a single qubit in the triangular lattice.  By extension, the ground states in the presence of a perturbative transverse field are matched, where the single qubit superposition $(\ket\uparrow + \ket\downarrow)/\sqrt 2$ is realized by a GHZ state over four qubits in the square octagonal case $(\ket{\uparrow\uparrow\uparrow\uparrow} + \ket{\downarrow\downarrow\downarrow\downarrow})/\sqrt 2$.

With this in mind one may expect that the triangular lattice can be used as a proxy for the square-octagonal lattice; this would require a mapping of $\Gamma/J$ and $T/J$ in square-octagonal lattice to $\Gamma_t/J$ and $T_t/J$ the triangular lattice such that the quantum critical points and finite-temperature phase transitions match.  It would also require that the support of the square-octagonal wavefunction be dominated by chain-intact states $\ket{\uparrow\uparrow\uparrow\uparrow}$, $\ket{\downarrow\downarrow\downarrow\downarrow}$, or their superposition.  However, even within the experimental parameters this is not the case, as evidenced by asymmetry between CCW and CW winding around the cylinder.  These directions are symmetric in the triangular lattice, but not in the square-octagonal lattice; this subtlety is accurately simulated at equilibrium by QA, as shown in Fig.~\ref{fig:winding_convergence}c.  The asymmetry is also clearly visible in the world-line dynamics of PIMC, as seen in convergence from CCW and CW initial conditions (Fig.~\ref{fig:2}b).

\subsubsection{Spin vector model}
The standard spin-vector rotor model associates angles $\theta \in [0,\pi]^N$ to qubits on the lattice, and has Hamiltonian
$$H(\theta) = \sum_{ij} J_{ij} \cos(\theta_i) \cos(\theta_j) + \sum_i h_i \cos(\theta_i) - \Gamma \sum_i \sin(\theta_i),$$
derived by replacing $\sigma^z$ and $\sigma^x$ in the TFIM Hamiltonian with $\cos(\theta)$ and $\sin(\theta)$ respectively.  A spin vector Monte Carlo (SVMC, also SSSV after proponents~\cite{Shin2014}) was proposed as a semiclassical model that approximately reproduces QA performance in certain situations.  Entanglement in a QA processor was later demonstrated experimentally \cite{Lanting2014} in agreement with the quantum adiabatic master equation and to the exclusion of SVMC \cite{Albash2015b}.
While the spin-vector ground state can be identified with the ground state of a quantum system that is separable at the level of qubits, we have already demonstrated the presence of entanglement both within and between four-qubit chains.  Since the ground state of a ferromagnetic chain in the rotor model under a perturbative transverse field is not aligned with the transverse field (this would be analogous to delocalized superposition of the chain), the rotor model cannot reproduce the low-temperature physics of the square-octagonal lattice.  Furthermore the square-octagonal lattice has no quantum phase transition in the spin vector model, so any mapping between SVMC and PIMC in this setting will be approximate at best.  As shown in Fig.~\ref{fig:sm_postprocess}c--d, SVMC deviates significantly from both QA and PIMC in the range of parameters studied.

\subsection{Additional observables}

\begin{figure}
  \includegraphics[scale=.7]{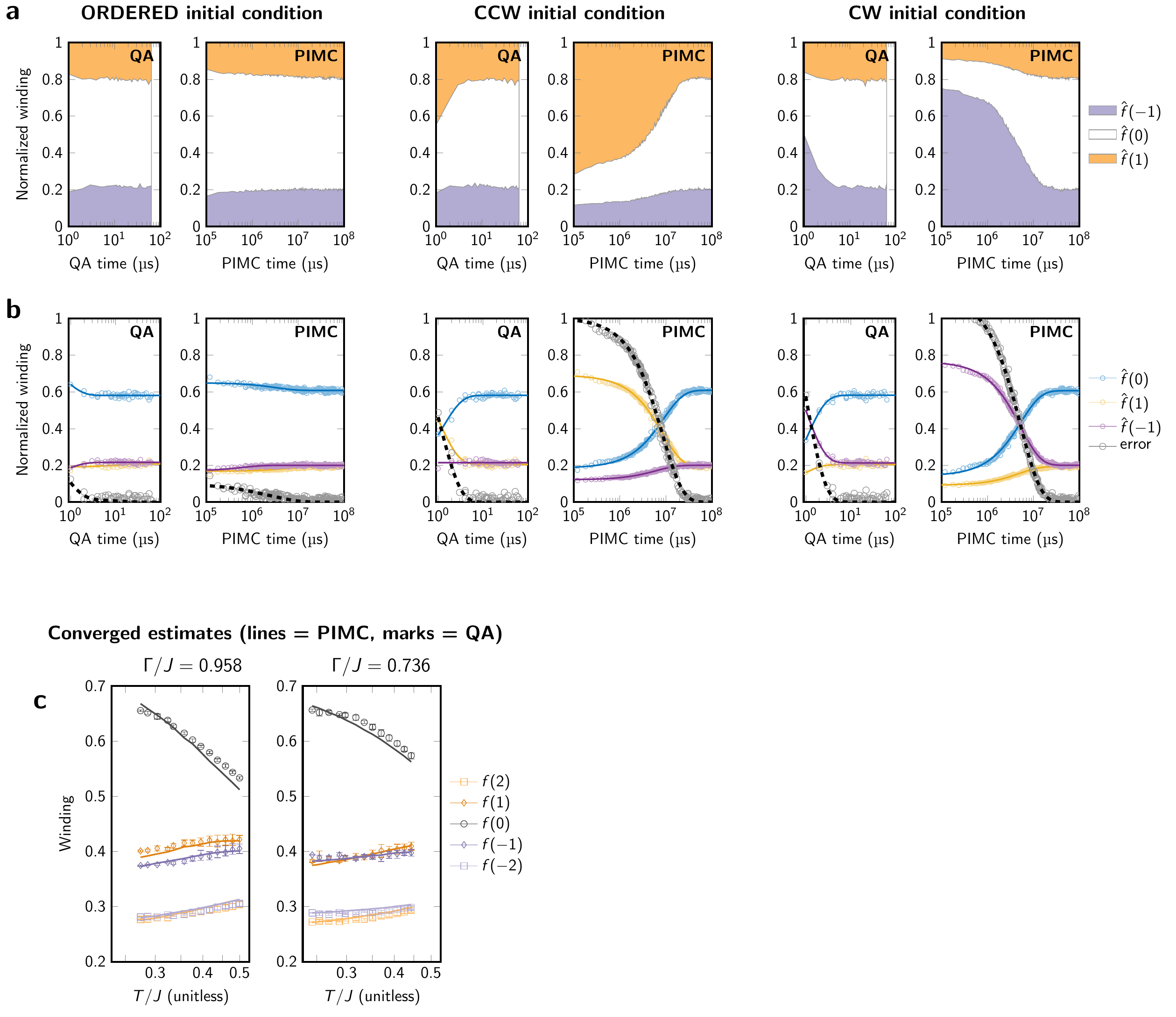}
  \caption{{\bf Convergence of winding number in QA and PIMC.}  Analogous to the data presented in Fig.~\ref{fig:2}b, we can observe convergence of winding number $f(w)$ from various initial conditions, in QA and PIMC (quenched).  Data shown is for $f(0)$, $f(1)$, and $f(-1)$, normalized by their sum to give $\hat f(w) = f(w)/(f(0)+f(1)+f(-1))$.  {\bf a}, Area plots of the three normalized winding numbers $\hat f(w)$ shows convergence to a stable distribution for each simulation regardless of initial condition. {\bf b}, Normalized winding numbers $\hat f(w)$ are plotted separately, along with exponential fit functions.  The sum of these fits' deviations from their equilibrium values gives a measure of error on the distribution, which converges to zero.  As in Fig.~\ref{fig:2}b, data shown correspond to $\Gamma/J=0.736$, $T/J = 0.244$.
  {\bf c}, Equilibrium estimates of non-normalized winding numbers $f(w)$ for QA (markers) and PIMC (lines) are shown at two values of $\Gamma/J$ corresponding to $s=0.36$ and $s=0.38$.  Beyond the broad quantitative agreement, QA accurately captures subtle equilibrium properties of the square-octagonal lattice, in particular the asymmetry between negative and positive winding and its dependence on $\Gamma$ and $T$.  Error bars are $95\%$ bootstrap confidence intervals.}\label{fig:winding_convergence}
\end{figure}

\begin{figure}
  \includegraphics[scale=.7]{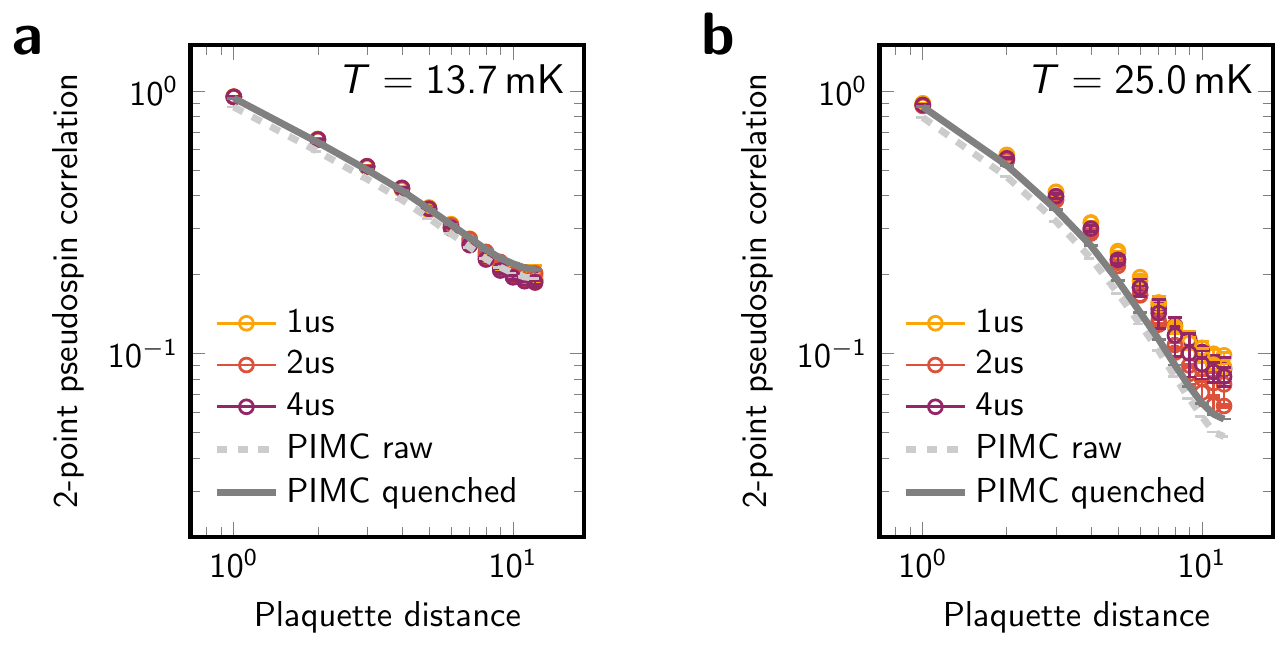}
  \caption{{\bf Two-point correlation functions.}  We measure two-point pseudospin correlations along the periodic axis of the cylinder for the $1440$-spin lattice and find good agreement between QA and PIMC.  Data shown are for $\Gamma/J = 0.736$ as in Fig.~\ref{fig:2}c--d.  Long-range correlations are overestimated for high temperature, consistent with the hypothesis of a limited global ordering in the lattice during the QA readout quench.
  }\label{fig:corr}
\end{figure}

\begin{figure}
  \includegraphics[scale=.7]{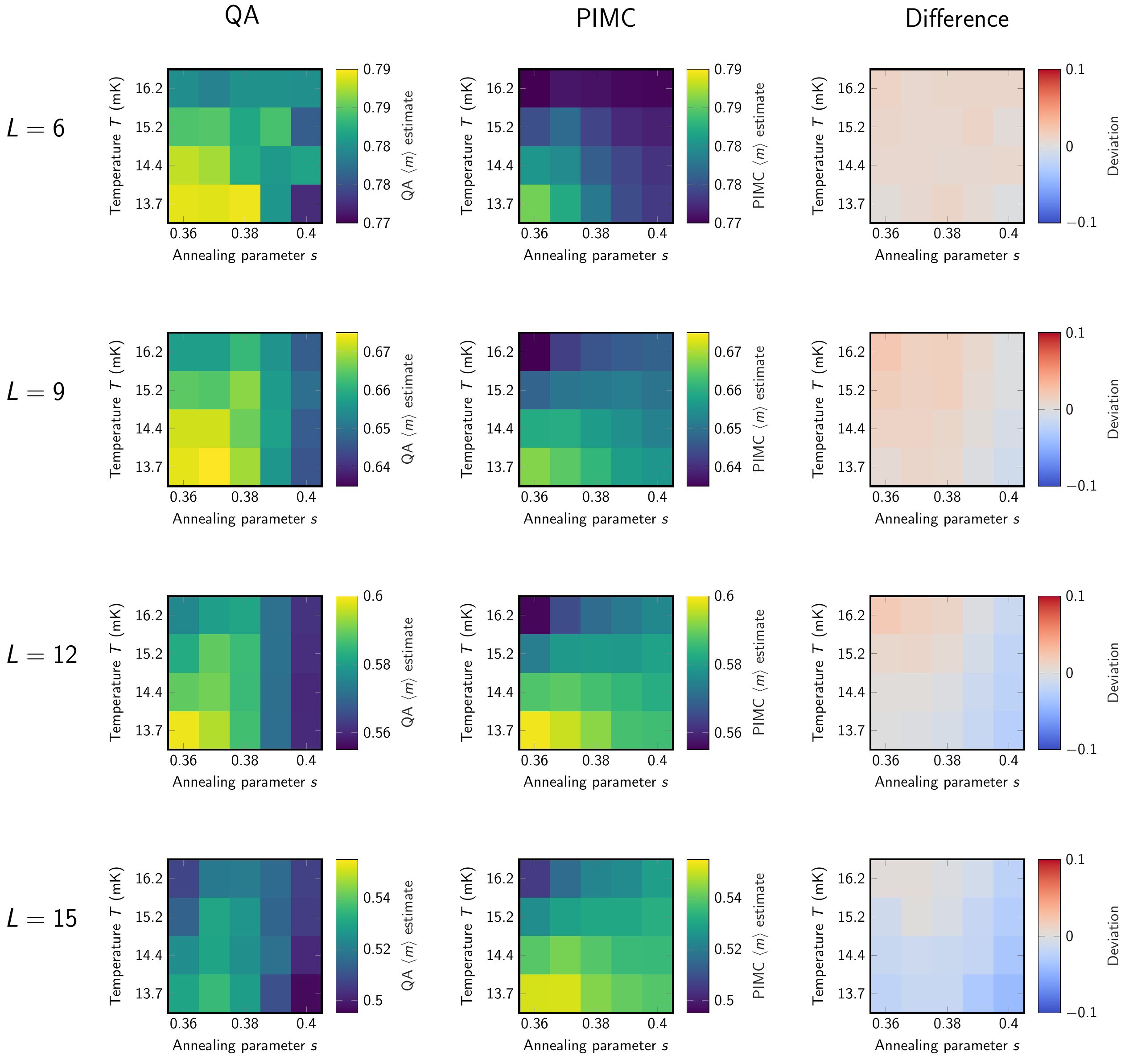}
  \caption{{\bf QA and PIMC equilibrium estimates of order parameter.} Analogous to Fig.~\ref{fig:3}b and c, we show data for all system sizes studied for the four coldest temperatures and slowest-converging annealing parameters studied.  These estimates are derived from an average of ordered and random initial conditions.}\label{fig:heatmaps}
\end{figure}

Here we provide data on additional observables.  Fig.~\ref{fig:winding_convergence} shows convergence and equilibrium values of winding number.  Both PIMC and QA exhibit asymmetry between counterclockwise and clockwise winding that varies as a function of $T/J$ and $\Gamma/J$.  This asymmetry is a subtle detail of equilibrium statistics in the square-octagonal lattice that is absent in the related triangular antiferromagnet, where there exists an additional flip symmetry of the cylinder.  It is important for two reasons.  First, it indicates that the quantum simulation is accurate enough to exhibit highly nontrivial physics, thereby demonstrating the utility of quantum simulation.  Second, it immediately rules out several simplified models of the lattice such as the triangular AFM, six-state clock model, and any other model from which this asymmetry is absent.

Fig.~\ref{fig:corr} shows two-point pseudospin correlations along the periodic axis of the cylinder, taken far from the open boundaries as in previous work\cite{King2018}.  The decay form of these correlations are important witnesses of critical phenomena\cite{Isakov2003,King2018}.

Fig.~\ref{fig:heatmaps} reproduces the data in Fig.~\ref{fig:3}b--c for smaller instances, showing accuracy in estimating $\braket m$ for every system size studied, usually to within $0.01$ of converged PIMC estimates.

\end{document}